\documentclass[12pt]{article}
\usepackage{epsfig,amsmath}
\def\beqn{\begin{eqnarray}}
\def\eeqn{\end{eqnarray}}
\def\vs{\vspace{0.3cm}}

\begin{document}
\pagestyle{empty}
\baselineskip=0.212in
\renewcommand{\theequation}{\arabic{section}.\arabic{equation}}
\renewcommand{\thefigure}{\arabic{figure}}
\renewcommand{\thetable}{\arabic{section}.\arabic{table}}

%%%%%%%%%%%%%%%%%%%%%%%%%%%%%%%%%%%%%%%%%%%%%%%%%%%%%%%%%%%%%%%%%%%%%%%%%%%%%%
%%%%%%%%%%%%%%%%%%%%%%%%%%%%%%%%%%%%%%%%%%%%%%%%%%%%%%%%%%%%%%%%%%%%%%%%%%%%%%
\begin{flushleft}
\large
{SAGA-HE-117-97 (hep-ph/9707220)
%%\hfill \today } \\
\hfill October 28, 1997}  \\
\end{flushleft}

\vspace{1.5cm}
 
\begin{center}
 
\Large{{\bf Numerical solution of $\bf Q^2$ evolution equations}} \\
\vspace{0.3cm}
\Large{{\bf for polarized structure functions}} \\

\vspace{1.3cm}
 
\Large
{M. Hirai, S. Kumano, and M. Miyama $^*$ }         \\
 
\vspace{0.5cm}
 
\Large
{Department of Physics, Saga University, Saga 840, Japan} \\

\vspace{1.5cm}

\Large{ABSTRACT}
  
\end{center}

We investigate numerical solution of 
Dokshitzer-Gribov-Lipatov-Altarelli-Parisi \\
(DGLAP) $Q^2$ evolution equations for 
longitudinally polarized structure functions.
Flavor nonsinglet and singlet equations with next-to-leading-order
$\alpha_s$ corrections are studied. A brute-force method is employed.
Dividing the variables $x$ and $Q^2$ into small steps, we simply solve
the integrodifferential equations.
Numerical results indicate that accuracy is better than 1\%
in the region $10^{-5}<x<0.8$ if more than two-hundred $Q^2$ steps
and more than one-thousand $x$ steps are taken.
Our evolution results are compared with
polarized experimental data of the spin asymmetry $A_1$ by
the SLAC-E130, SLAC-E143, EMC, and SMC collaborations.
The comparison indicates that we cannot assume $A_1$ is
independent of $Q^2$.
We provide a FORTRAN program for the Q$^2$ evolution and devolution
of polarized nonsinglet-quark, 
singlet-quark, $\Delta q_i+\Delta\bar q_i$, and gluon distributions 
(and corresponding structure functions).

\vspace{0.8cm}

\vfill
 
\noindent
{\rule{6.cm}{0.1mm}} \\
 
\vspace{-0.3cm}
\normalsize
\noindent
* Email: 96sm18@edu.cc.saga-u.ac.jp, 
          kumanos@cc.saga-u.ac.jp, and \\

\vspace{-0.6cm}
\noindent
96td25@edu.cc.saga-u.ac.jp. Information on their research is available \\

\vspace{-0.6cm}
\noindent
at http://www.cc.saga-u.ac.jp/saga-u/riko/physics/quantum1/structure.html. \\

\vspace{+0.1cm}
\hfill
{published as Comput. Phys. Commun. 108 (1998) 38.}

\normalsize

%\vfill\eject
%%%%%%%%%%%%%%%%%%%%%%%%%%%%%%%%%%%%%%%%%%%%%%%%%%%%%%%%%%%%%%%%%%%%%%%%%%%%%%
%%%%%%%%%%%%%%%%%%%%%%%%%%%%%%%%%%%%%%%%%%%%%%%%%%%%%%%%%%%%%%%%%%%%%%%%%%%%%%
%\pagestyle{empty}
%\tableofcontents
%\addtocontents{toc}{\protect\vspace{0.7cm}}
%%%%%%%%%%%%%%%%%%%%%%%%%%%%%%%%%%%%%%%%%%%%%%%%%%%%%%%%%%%%%%%%%%%%%%%%%%%%%%
%%%%%%%%%%%%%%%%%%%%%%%%%%%%%%%%%%%%%%%%%%%%%%%%%%%%%%%%%%%%%%%%%%%%%%%%%%%%%%

\vfill\eject
%%%%%%%%%%%%%%%%%%%%%%%%%%%%%%%%%%%%%%%%%%%%%%%%%%%%%%%%%%%%%%%%%%%%%%%%%%%%%%
%%%%%%%%%%%%%%%%%%%%%%%%%%%%%%%%%%%%%%%%%%%%%%%%%%%%%%%%%%%%%%%%%%%%%%%%%%%%%%
\pagestyle{plain}
\setcounter{page}{1}

\section*{{\bf Program Summary}}
\addcontentsline{toc}{section}{\protect\numberline{\S}{Program Summary}}

\ \\
\noindent
{\it Title of program:} BFP1

\vs
\noindent
{\it Computer:} AlphaServer 2100 4/200; 
               {\it Installation:} 
                     The Research Center for Nuclear Physics in Osaka

\vs
\noindent
{\it Operating system:} OpenVMS V6.1

\vs
\noindent
{\it Programming language used:} FORTRAN 77

\vs
\noindent
{\it Peripherals used:} Laser printer

\vs
\noindent
{\it No. of lines in distributed program, including test data, etc.:} 
1617

\vs
\noindent
{\it Keywords:} Structure function, polarized parton distribution, 
                Q$^2$ evolution, numerical solution.

\vs
\noindent
{\it Nature of physical problem} 

\noindent
This program solves DGLAP $Q^2$ evolution equations
with or without next-to-leading-order $\alpha_s$ effects
for longitudinally polarized parton distributions.
The evolved distributions could be convoluted
with coefficient functions for calculating
the structure function $g_1$.
Both flavor-nonsinglet and singlet cases are provided, so that
the distributions,
$x\Delta q_{_{NS}}$, $x\Delta q_{_S}$, 
$x\Delta q_i^+\equiv x\Delta q_i+x\Delta \bar q_i$ ($i$=quark flavor),
$x\Delta g$, $xg_{1_{NS}}$, $xg_{1_{S}}$, and $xg_{1,i}^+$ 
can be obtained.

\vs
\noindent
{\it Method of solution}

\noindent
The DGLAP integrodifferential equations are simply solved
by a brute-force method.
We divide the variables $x$ and $Q^2$ into very small steps,
and the integrodifferential equations are solved step by step.

\vs
\noindent
{\it Restrictions of the program}

\noindent
This program is used for calculating Q$^2$ evolution of 
a longitudinally polarized
flavor-nonsinglet-quark, singlet-quark, $\Delta q_i^+$, 
and gluon distributions 
in the leading order or in the next-to-leading order of $\alpha_s$.
$Q^2$ evolution equations are the DGLAP equations.
The double precision arithmetic is used.
The renormalization scheme is the modified minimal subtraction
scheme ($\overline{MS}$).
A user provides the initial distribution as a subroutine
or as a data file.
Examples are explained in section \ref{INPUT}.
Then, the user inputs fifteen parameters in section \ref{INPUT}.

\vs
\noindent
{\it Typical running time}

\noindent
Approximately seven minutes on AlphaServer 2100 4/200 in the nonsinglet
case, sixty minutes in the singlet-quark evolution.

\vfill\eject
%%%%%%%%%%%%%%%%%%%%%%%%%%%%%%%%%%%%%%%%%%%%%%%%%%%%%%%%%%%%%%%%%%%%%%%%%%%%%%
%%%%%%%%%%%%%%%%%%%%%%%%%%%%%%%%%%%%%%%%%%%%%%%%%%%%%%%%%%%%%%%%%%%%%%%%%%%%%%
\section*{{\bf LONG WRITE-UP}}

\noindent
\section{{\bf Introduction}}\label{INTRO}
\setcounter{equation}{0}
\setcounter{figure}{0}
\setcounter{table}{0}

Spin structure of the proton has been investigated by polarized
lepton-proton scattering.  In particular, the structure function
$g_1$ provides us information on the probability to find a quark
polarized along the direction of longitudinally polarized
proton spin minus the one polarized oppositely. Experimentally, 
it is measured by a spin asymmetry in the lepton-proton scattering. 
The $g_1$ is a function of two kinematical variables $Q^2$ 
and $x=Q^2/(2p\cdot q)$, where $p$ is the proton momentum
and $Q^2$ is given by the four-momentum transfer $q$ as $Q^2=-q^2$.
The $x$ dependence is associated with nonperturbative physics, so 
that $x$ distributions of polarized parton distributions are obtained
by fitting various experimental data.
At this stage, the only way to study the parton $x$ distributions
theoretically is to use phenomenological quark models. On the other hand,
the $Q^2$ dependence is well predicted in perturbative QCD. 
In general, structure functions are almost independent of the scale $Q^2$,
which is called Bjorken scaling. However, even though it is small,
scaling-violation phenomena are observed. The experimental data support
perturbative QCD predictions.

As a way of describing the scaling violation, 
Dokshitzer-Gribov-Lipatov-Altarelli-Parisi (DGLAP) equations \cite{AP} are 
usually used. They are coupled integrodifferential equations.
Because the DGLAP equations are frequently used in theoretical and
experimental studies, it is worth while creating a program to solve
them accurately. We have been studying this topic last several years
in a Laguerre-polynomial method \cite{LAG} and in a brute-force method
\cite{BF,BF2}. Here, we extend the studies to the spin-dependent case.
Fortunately, the next-to-leading-order (NLO) splitting functions are
calculated recently for polarized parton distributions \cite{PNLO}. 
It enables us to investigate the details of NLO effects on the $Q^2$
evolution. We create a FORTRAN program for obtaining numerical solution
of the polarized DGLAP equations with the NLO effects. 
With our program, $Q^2$ evolution of nonsinglet quark, singlet quark,
and gluon distributions can be obtained.
Furthermore, NLO coefficient functions could be convoluted with
the evolved distributions in our program
for calculating the $g_1$ structure function.
It is very useful for investigating polarized parton distributions.
In fact, our program in a slightly modified form is used for
investigating optimum polarized parton distributions by fitting
experimental data \cite{RHIC-J}.

We explain the details of our studies in the following.
In section \ref{DGLAP}, the polarized DGLAP $Q^2$ evolution equations
are explained. Then, our numerical solution method is described
in section \ref{BFMETHOD}.
Input parameters and input distributions are discussed in 
section \ref{INPUT}, and essential subroutines in the program are 
explained in section \ref{BFP1}.
Numerical results and their comparison with experimental data
are discussed in section \ref{RESULTS}. 
Summary is given in section \ref{SUMMARY}.
Explicit forms of splitting functions and the coefficient functions
are listed in Appendices.

\vfill\eject
%%%%%%%%%%%%%%%%%%%%%%%%%%%%%%%%%%%%%%%%%%%%%%%%%%%%%%%%%%%%%%%%%%%%%%%%%%%%%%
%%%%%%%%%%%%%%%%%%%%%%%%%%%%%%%%%%%%%%%%%%%%%%%%%%%%%%%%%%%%%%%%%%%%%%%%%%%%%%
\section{{\bf $\bf Q^2$ evolution equations}}\label{DGLAP}
\setcounter{equation}{0}
\setcounter{figure}{0}
\setcounter{table}{0}

We use the DGLAP equations \cite{AP} for studying the $Q^2$ evolution.
Both the leading order (LO) and the next-to-leading 
order (NLO) cases can be handled by the DGLAP equations.
NLO effects are included in the running coupling constant 
$\alpha _s (Q^2)$ and in the splitting functions $\Delta P_{ij} (x)$. 
Here, the evolution of polarized parton distributions 
$\Delta q = q_{\uparrow}-q_{\downarrow}$ and
$\Delta g = g_{+1}-g_{-1}$ is investigated.

First, the nonsinglet DGLAP equation is given by 
\beqn
\frac{\partial}{\partial\ln Q^2} \, \Delta q_{_{NS}}(x,Q^2)\ = 
\frac{\alpha_s (Q^2)}{2\pi} \ 
\Delta P_{q^\pm, {NS}} (x) \otimes \Delta q_{_{NS}}(x,Q^2) 
\ ,
\label{nonap}
\eeqn
where $\Delta q_{_{NS}}(x,Q^2)$ is a polarized nonsinglet parton distribution, 
$\Delta P_{q^\pm, {NS}}$ is the polarized nonsinglet splitting function, and
the convolution $\otimes$ is defined by
\beqn
f (x) \otimes g (x) = \int^{1}_{x} \frac{dy}{y} f\left(\frac{x}{y} \right) g(y) 
\ .
\eeqn
The notation $q^\pm$ in the splitting function indicates
a $\Delta q^+ = \Delta q + \Delta \bar q$ or 
  $\Delta q^- = \Delta q - \Delta \bar q$ distribution type, which is
explained in Appendix A.
Instead of $Q^2$, it is more convenient to use the variable $t$ defined by
\beqn
t \equiv - \frac{2}{\beta_0} \ln 
\left[ \frac{\alpha_s(Q^2)}{\alpha_s(Q_0^2)} \right] \ ,
\label{t}
\eeqn
where $\beta_0$ is defined in Appendix A.
The parton distribution and the splitting function multiplied by $x$
\beqn
\widetilde f (x) = x f(x) 
\eeqn
satisfy the same integrodifferential equation.
Therefore, we rewrite the evolution equation as
\beqn
\frac{\partial}{\partial t} \ \Delta \widetilde q_{_{NS}} (x,t) =  
\Delta \widetilde P_{q^\pm, {NS}} (x) \otimes \Delta \widetilde q_{_{NS}}(x,t)
\ .
\label{nonsiglet}
\eeqn

Next, the singlet evolution is more complicated than the nonsinglet one
due to gluon participation in the evolution. The singlet quark distribution
is defined by 
$\Delta \widetilde{q}_s (x,t) \equiv \sum_i^{N_{f}} x\, \Delta q_i^+ $ 
where $i$ is the flavor, 
and $\Delta \widetilde g(x,t) $ is the gluon distribution.
The singlet case is given by 
\beqn
\frac{\partial}{\partial t}   
\left(\begin{array}{c} 
  \Delta \widetilde{q}_s (x,t) \\ 
  \Delta \widetilde g (x,t)  
\end{array} \right) &=&
\left( \begin{array}{cc} 
  \Delta \widetilde P_{qq}(x,t) & \Delta \widetilde P_{qg}(x,t) \\  
  \Delta \widetilde P_{gq}(x,t) & \Delta \widetilde P_{gg}(x,t) \\ 
\end{array} \right) \otimes
\left( \begin{array}{c}
  \Delta \widetilde{q}_s (x,t) \\
  \Delta \widetilde g(x,t)
\end{array} \right) 
\ .
\label{singlet}
\eeqn
If each flavor evolution is necessary,
another equation has to be used in addition to Eq. (\ref{singlet}).
Using the properties of the splitting functions, 
$\Delta \widetilde{P}^{(1)}_{q^+_{i} q^+_j} = 
  \delta_{ij} \, \Delta \widetilde{P}^{(1)}_{q^+, _{NS}} 
  + 2C_F T_R \widetilde{F}_{qq}$
and $\Delta \widetilde{P}^{(1)}_{q^+_j g} = 
     \Delta \widetilde{P}^{(1)}_{q^+ g}$ (independent of $j$),
we have the evolution equation
\begin{align}
\frac{\partial}{\partial t} \Delta \widetilde{q}_i^{\, +} (x,t) = 
\Delta \widetilde{P}_{q^+,\, _{NS}} (x) \otimes 
\Delta \widetilde q_i^{\, +} (x,t)
&+ \ 2 \, C_F \, T_R \, 
  \widetilde F_{qq} (x) \otimes \Delta \widetilde q_{_S} (x,t) 
\nonumber \\ 
&+  \Delta \widetilde{P}_{q_i^+ g} (x) \otimes 
               \Delta \widetilde g (x,t)
\ .
\label{flavorap} 
\end{align}
In this case, three coupled integrodifferential equations
in Eqs. (\ref{singlet}) and (\ref{flavorap}) should be solved simultaneously.
All the necessary splitting functions $\Delta \tilde P$ and $\tilde F$
are listed in Appendix A.

We discuss the NLO effects in the evolution equations. 
The NLO contributions are included 
in the running coupling constant $\alpha _s(t)$, in the splitting functions 
$\Delta P_{ij}(z)$, and in the coefficient functions. Once the NLO corrections 
are included in the evolution, the renormalization scheme has to be specified. 
We use the $\overline{MS}$ scheme throughout this paper.

The running coupling constant in the leading order (LO) is given by
\beqn
\alpha_s^{LO}(Q^2) = \frac{4\pi}{\beta_0 \ln (Q^2/\Lambda^2)} 
\ ,
\eeqn
and the one in the next-to-leading order (NLO) is 
\beqn
\alpha_s^{NLO}(Q^2)=\frac{4\pi}{\beta_0 \ln(Q^2/\Lambda^2)}
\left[1-\frac{\beta_1 \ln \{ \ln(Q^2/\Lambda^2) \} }
{\beta_0^2\ln(Q^2/\Lambda^2)}
\right]  
\ .
\eeqn
The constants $\beta_0$ and $\beta_1$ are defined in Appendix A.
The splitting functions have the perturbative-expansion form
\beqn
\Delta P_{ij}(x) = 
\Delta P_{ij}^{(0)}(x) + \frac{\alpha_s(Q^2)}{2\pi} \,
\Delta P_{ij}^{(1)}(x)
\ .
\label{dpij}
\eeqn
The second term is the NLO contributions to the splitting functions.
The functions are expressed by the ones multiplied by $x$ 
($\Delta \widetilde P$).
Changing the variable $Q^2$ to $t$ in the DGLAP equations, we have
the splitting functions 
\beqn
\Delta \widetilde P_{ij}(x) = \Delta \widetilde P_{ij}^{(0)}(x)
                +\frac{\alpha_s(t)}{2\pi} \Delta R_{ij}(x)
\ ,
\label{dptilij}
\eeqn
where the function $\Delta R_{ij}(x)$ is
\beqn
\Delta R_{ij} (x)\equiv \Delta \widetilde {P}_{ij}^{(1)}(x) 
- \frac{\beta_1}{2\beta_0} \, \Delta \widetilde {P}_{ij}^{(0)}(x)  
\ .
\label{Rij}
\eeqn
The second term in Eq. (\ref{Rij})
appears because of the transformation from $Q^2$ to $t$.
To be precise, the splitting functions 
$\Delta \widetilde P_{ij}$ in Eq. (\ref{dptilij}) should be denoted,
for example, $\Delta \widetilde P_{ij}'$ because it is different from 
$x \Delta P_{ij}=x \Delta P_{ij}^{(0)}+(\alpha_s/2\pi)x \Delta P_{ij}^{(1)}$
which is Eq. (\ref{dpij}) multiplied by $x$.
(Note the definition of $\tilde f$ is $\tilde f=x f$.)
However, we use the expression without the prime
throughout this paper for simplicity.
We have the same expression 
$\Delta \widetilde P_{NS}= \Delta \widetilde P^{(0)}_{NS}
  + \alpha_s/(2\pi) \Delta R_{NS}$
in the nonsinglet case.

Next, we discuss the spin-dependent structure function $g_1$.
We have discussed how the $Q^2$ evolution of quark and gluon distributions
is described. However, these parton distributions are not observed directly
in experiments. The $g_1$ could be measured in the polarized lepton-proton
scattering with information on the unpolarized structure function $F_1$.
In the LO case, it is given in parton model as
\beqn
x g_1(x) = \frac{1}{2} \sum_i^{N_f} e^2_i \, 
             \Delta \widetilde q_i^{\, +} (x)
\ .
\label{g1lo}
\eeqn 
The NLO effects in the structure function are included in the coefficient
functions and also in the quark and gluon distributions.
In the NLO case, the quark distributions should be convoluted with
a coefficient function. Furthermore, 
an additional gluon correction term should be taken into account:
\beqn
x g_1(x) = \frac{1}{2} \sum_i^{N_f} e^2_i \,  
\left[ \, \Delta \widetilde C_q (x) \otimes \Delta {\widetilde q}_i^{\, +} (x) 
     +  \Delta \widetilde C_g (x) \otimes \Delta \widetilde g (x) 
\, \right]
\ , 
\label{g1nlo}
\eeqn
where $\Delta \widetilde C_q/x$ and
$\Delta \widetilde C_g/x$ are the quark and gluon coefficient functions
in Appendix B.
The calculation procedure for the $g_1$ evolution is the following.
First, the $Q^2$ evolution of the quark and gluon distributions is calculated. 
Then, the structure function $g_1$ at $Q^2$ is evaluated by using 
Eq. (\ref{g1lo}) or (\ref{g1nlo}).
An example is explained in section \ref{makeg1}.

\vfill\eject
%%%%%%%%%%%%%%%%%%%%%%%%%%%%%%%%%%%%%%%%%%%%%%%%%%%%%%%%%%%%%%%%%%%%%%%%%%%%%%
%%%%%%%%%%%%%%%%%%%%%%%%%%%%%%%%%%%%%%%%%%%%%%%%%%%%%%%%%%%%%%%%%%%%%%%%%%%%%%
\section{{\bf Brute-force method}}\label{BFMETHOD}
\setcounter{equation}{0}
\setcounter{figure}{0}
\setcounter{table}{0}

Among various methods of solving the DGLAP equations,
we decide to employ a brute-force method.
So far, we have investigated two methods, the Laguerre-polynomial \cite{LAG}
and the brute-force \cite{BF} methods, in the spin-independent case.
The Laguerre method has an advantage of computing time. However, numerical
accuracy becomes slightly worse in the nonsinglet case at small $x$.
As far as we studied, the situation is better in the polarized distributions.
However, we find a tendency that the accuracy is slightly worse at small $x$.
In light of future HERA spin physics, our program should be accurately
enough even at $x=10^{-5}$. Therefore, we consider that the brute-force
method is safer for getting accurate results in the wide $x$ range.
The variables $x$ and $t$ are divided into small steps, then
integration and differentiation are defined by
\begin{align}
\frac{d f(x)}{dx} &= \frac{f(x_{m+1})-f(x_m)}{\delta x_m} \ ,\\
\int f(x)\ dx &= \sum_{m=1}^{N_x} \delta x_m f(x_m) \ .
\end{align}
With these replacements, the evolution equations could be solved rather easily.
For example, Eq. (\ref{nonsiglet}) is written in the following form:
\beqn
\Delta \widetilde{q}_{_{NS}}(x_k,t_{j+1}) = 
  \Delta \widetilde{q}_{_{NS}}(x_k,t_j) + 
  \delta t_j \, \sum_{m=k}^{N_x} \frac{\delta x_m}{x_m}
\, \Delta \widetilde{P}_{q^\pm, NS} \left( \frac{x_k}{x_m} \right)
\, \Delta \widetilde{q}_{_{NS}}(x_m,t_j)
\ .
\label{bfnon}
\eeqn
First, the evolution from $t_0=0$ to $t_1=\delta t$ is calculated
in the above equation by providing the initial distribution 
${\Delta \widetilde{q}}_{_{NS}}({x}_{m},{t}_{j=1})$.
Repeating this step $N_t-1$ times, we obtain the final distribution
at $t_{N_t}$. Accurate results cannot be obtained at small $x$
if the linear $x$ step is taken ($\delta x=1/N_x$).
Therefore, the logarithmic-$x$ step 
$\delta (log_{10} x)=|log_{10} x_{min}|/N_x$ 
is taken in our evolution calculations.

The same method can be applied to the singlet and $\Delta q_i^+$ evolution
equations in Eqs. (\ref{singlet}) and (\ref{flavorap}).
These equations are written in the brute-force method as:
\addtocounter{equation}{1}
\renewcommand{\theequation}{\arabic{section}.\arabic{equation}a}
\addtocounter{equation}{-1}
\begin{align}
\Delta \widetilde{q}_i^{\ +}(x_k,t_{j+1}) =& 
  \Delta \widetilde{q}_i^{\ +} (x_k,t_j) \, +   
\, \delta t_j \, \sum_{m=k}^{N_x} \frac{\delta x_m}{x_m}
\, \Delta \widetilde{P}_{q^+,NS} \left(\frac{x_k}{x_m} \right)
\, \Delta \widetilde{q}_i^{\, +}(x_m,t_j) 
\nonumber \\ & +
\, \delta t_j \, \sum_{m=k}^{N_x} \frac{\delta x_m}{x_m}
\, 2 \, C_F \, T_R \, \widetilde{F}_{qq} \left(\frac{x_k}{x_m} \right)
\, \Delta \widetilde{q}_{_{S}}(x_m,t_j)
\nonumber \\ & +
\, \delta t_j \, \sum_{m=k}^{N_x} \frac{\delta x_m}{x_m}
\, \Delta \widetilde{P}_{q_i^+ g}\left(\frac{x_k}{x_m} \right)
\, \Delta \widetilde{g}(x_m,t_j)
\ ,
\label{bfpi}
\end{align}
%\vspace{2cm}
\renewcommand{\theequation}{\arabic{section}.\arabic{equation}b}
\addtocounter{equation}{-1}
\begin{align}
\Delta \widetilde{q}_{_{S}}(x_k,t_{j+1}) =&
  \Delta \widetilde{q}_{_{S}}(x_k,t_j)\, +  
\, \delta t_j \, \sum_{m=k}^{N_x}\frac{\delta x_m}{x_m}
\, \Delta \widetilde{P}_{qq}\left(\frac{x_k}{x_m} \right) 
\, \Delta \widetilde{q}_{_{S}}(x_m,t_j) 
\nonumber \\ & +
\, \delta t_j \, \sum_{m=k}^{N_x} \frac{\delta x_m}{x_m}
\, \Delta \widetilde{P}_{qg}\left(\frac{x_k}{x_m} \right)
\, \Delta \widetilde{g}(x_m,t_j)
\ ,
\label{bfps}
\end{align}
\renewcommand{\theequation}{\arabic{section}.\arabic{equation}c}
\addtocounter{equation}{-1}
\begin{align}
\Delta \widetilde{g}(x_k,t_{j+1}) =&
  \Delta{\widetilde{g}}({x}_{k},{t}_{j})\ + 
\ \delta t_j \, \sum_{m=k}^{N_x} \frac{\delta x_m}{x_m}
\ \Delta \widetilde{P}_{gq}\left(\frac{x_k}{x_m} \right)
\ \Delta \widetilde{q}_{_{S}}(x_m,t_j) 
\nonumber \\ & +
\ \delta t_j \, \sum_{m=k}^{N_x} \frac{\delta x_m}{x_m}
\ \Delta \widetilde{P}_{gg}\left(\frac{x_k}{x_m} \right)
\ \Delta \widetilde{g}(x_m,t_j)
\ .
\label{bfpg}
\end{align}
\renewcommand{\theequation}{\arabic{section}.\arabic{equation}}

\noindent
These are coupled equations. However, they can be solved 
by providing the initial distributions,
$\Delta \widetilde{q}_i^{\, +} (x_m,t_{j=0})\ ,
\ \Delta \widetilde{q}_{_{S}}(x_m,t_{j=0})$,
and 
$\Delta \widetilde{g} (x_m,t_{j=0})$,
and by repeating the evolution step $N_t$ times.

We should be careful in handling 
$1/(1-x)_+$ terms in the splitting functions.
They are given in our method by
\beqn
\int_x^1 dx' \frac{f(x')}{(1-x')_+} = 
\sum_{m=k}^{N_x} \delta x_m \, \frac{f(x_m)-f(1)}{1-x_m}
\ + \ f(1) \, \ln (1-x_k) 
\ . 
\eeqn
In the same way, the integral with $[\ln (1-x)/(1-x)]_+$ becomes
\begin{align}
\int_x^1 dx' f(x') \left[ \frac{\ln (1-x')}{1-x'} \right]_+ = &
\sum\limits_{m=k}^{N_x} \delta x_m \, 
 [f(x_m)-f(1)]  \, \frac{\ln (1-x_m)}{1-x_m} 
\nonumber \\
& 
+ \frac{1}{2} \, f(1) \, \ln ^2 (1-x_k)
\ .
\end{align}

\vfill\eject
%%%%%%%%%%%%%%%%%%%%%%%%%%%%%%%%%%%%%%%%%%%%%%%%%%%%%%%%%%%%%%%%%%%%%%%%%%%%%%
%%%%%%%%%%%%%%%%%%%%%%%%%%%%%%%%%%%%%%%%%%%%%%%%%%%%%%%%%%%%%%%%%%%%%%%%%%%%%%
\section{{\bf Description of input parameters and input \\ distribution}}
\label{INPUT}
\setcounter{equation}{0}
\setcounter{figure}{0}
\setcounter{table}{0}

For running the FORTRAN-77 program BFP1, a user should supply
fifteen input parameters from the file \#10.
In addition, an input distribution(s) should be
given in a function subroutine(s) in the end of the FORTRAN program
or in an input data file(s), \#13, \#14, and/or \#15.
The initial distribution(s) could be written in the output file \#12.
Evolution results are written in the output file \#11.
We explain the input parameters and the input distributions
in the following.

%%%%%%%%%%%%%%%%%%%%%%%%%%%%%%%%%%%%%%%%%%%%%%%%%%%%%%%%%%%%%%%%%%%%%%%%%%%%%%
\subsection{Input parameters}\label{PARAMET}

There are fifteen parameters.
Numerical values of the parameters should be supplied
in the file \#10, then these are read
in the main program.

\vspace{0.3cm}

\small
\noindent
\begin{tabular}{|l|c|l|} \hline
IOUT   & 1 & write $x$ and $x\Delta q_{_{NS}}(x)$ [or $xg_{1_{NS}}(x)$]
              at fixed $Q^2$ (=Q2) in the file \#11
\\ \cline{2-3}
       &  2 & write $Q^2$ and $x\Delta q_{_{NS}}(Q^2)$ [or $xg_{1_{NS}}(Q^2)$]
              at fixed $x$ (=XX) in the file \#11
\\ \cline{2-3}
       & 3 & $x$, $x\Delta q_{_S}(x)$ [or $xg_{1_S}(x)$], and $x\Delta g(x)$ \\ 
                     \cline{2-3}
       & 4 & $Q^2$, $x\Delta q_{_S}(Q^2)$ [or $xg_{1_S}(Q^2)$],
             and $x\Delta g(Q^2)$
\\ \cline{2-3}
       & 5 & $x$, $x\Delta q_i^+(x)$ [or $xg_{1,i}^+(x)$],
             $x\Delta q_{_S}(x)$ [or $xg_{1_S}(x)$], 
             and $x\Delta g(x)$ [or $xg_{1,g}(x)$]
\\ \cline{2-3}
       & 6 & $Q^2$, $x\Delta q_i^+(Q^2)$ [or $xg_{1,i}^+(Q^2)$],
             $x\Delta q_{_S}(Q^2)$ [or $xg_{1_S} (Q^2)$], 
\\ 
       &   & and $x\Delta g(Q^2)$ [or $xg_{1,g}(x)$]
\\ \hline \hline
IREAD    & 1 & give initial distribution(s) in function subroutine(s)
\\ \cline{2-3}
         & 2 & read initial distribution(s) from data file(s)
\\ \hline \hline
INDIST   & 1 & do not write initial distribution(s)
\\ \cline{2-3}
         & 2 & write initial distribution(s) in the file \#12
\\ \cline{2-3}
         & 3 & write initial distribution(s) in the file \#12
               without calculating evolution
\\ \hline \hline
IORDER   & 1 & leading order (LO) in $\alpha_ s$
\\ \cline{2-3}
         & 2 & next-to-leading order (NLO)
\\ \hline \hline
ITYPE    & 1 & structure function $xg_1(x,Q^2)$
\\ \cline{2-3}
         & 2 & quark distribution $x\Delta q(x,Q^2)$
               and gluon distribution $x\Delta g(x,Q^2)$
\\ \hline\hline
IMORP    & 1 & $x\Delta q_i- x\Delta \bar q_i$ type distribution
\\ \cline{2-3}
         & 2 & $x\Delta q_i+ x\Delta \bar q_i$ type distribution
\\ \hline
\end{tabular}

%\vspace{5mm}

\noindent
\begin{tabular}{|l|l|} \hline
Q02   &  initial $Q^2$ ($\equiv Q_0^2$ in GeV$^2$ )
                   at which an initial distribution is supplied
\\ \hline
Q2    &  $Q^2$ to which the distribution is evolved 
         ($Q^2\ne Q_0^2$, $Q^2 > Q_0^2$ or $Q^2 < Q_0^2$)
\\ \hline
DLAM  &  QCD scale parameter $\Lambda_{QCD}$ in GeV
\\ \hline
NF    &   number of quark flavors 
\\ \hline
XX    &  $x$ at which $Q^{2}$ dependent distributions are written
(IOUT=2, 4, or 6 case)
\\ \hline
NX    &  number of $x$ steps (NX$<$5000)
\\ \hline
NT    &  number of $t$ steps (NT$<$5000)
\\ \hline
NSTEP &  number of $x$ steps or $t$ steps for writing output distribution(s) 
\\ \hline
NXMIN & $log_{10}$(minimum of $x$) $[0<min(x)= 10^{^{NXMIN}}<XX]$
\\ \hline
\end{tabular}
\normalsize

\vspace{0.5cm}

The meaning of IREAD is explained in section 4.2.
The structure function $xg_1^{p}$ is obtained by the convolution
of the distributions, 
$(1/2)[(4/9)x (\Delta u+ \Delta \bar u + \Delta c + \Delta \bar c)+
(1/9)x (\Delta d+\Delta \bar d+\Delta s+\Delta \bar s)]$ and $x\Delta g$
in the four flavor case, with the corresponding coefficient functions.
Practically, we use the expression 
$(1/2)\sum_i e_i^2 x \Delta q_i=(1/2) x [4\Delta q_s
          - 3(\Delta d^+ +\Delta s^+)]/9$,
where $\Delta d^+ = \Delta d+\Delta \overline d$
and $\Delta s^+ = \Delta s+\Delta \overline s$,
instead of the above one in calculating $x g_1$.
It is explained in more detail in section \ref{makeg1}.
The expression could be used only in the four and three flavor cases.
In the five or six flavor evolution, it should be slightly modified.
The parameter IMORP indicates a plus or minus type distribution
$\sum_i a_i x (\Delta q_i \pm \Delta \bar q_i)$, where $a_i$ are some
constants. IMORP=1 or 2 should be taken in the nonsinglet evolution.
In the singlet or $\Delta q_i^+$ case, IMORP=2 should be chosen.

For example, if one would like to evolve an initial
singlet-quark distribution $x\Delta q_{_S}$ at $Q^2$=4 GeV$^2$
to the distribution at $Q^2$=200 GeV$^2$
by the NLO DGLAP equations with $N_f$=4 and $\Lambda$=0.231 GeV,
the input parameters could be
IOUT=3, IREAD=1, INDIST=1, IORDER=2, ITYPE=2, IMORP=2, 
Q02=4.0, Q2=200.0, DLAM=0.231, NF=4, XX=0.0,
NX=1000, NT=200, NSTEP=100, and NXMIN=$-$4.
In this case,  the input file \#10 is the following:

\vspace{0.3cm}
\noindent
\ \ \ \ \ \ 3, 1, 1, 2, 2, 2 

\noindent
\ \ \ \ \ \ 4.0, 200.0, 0.231, 4, 0.0, 1000, 200, 100, $-$4 \ .

%%%%%%%%%%%%%%%%%%%%%%%%%%%%%%%%%%%%%%%%%%%%%%%%%%%%%%%%%%%%%%%%%%%%%%%%%%%%%%
\subsection{Input distributions supplied by function subroutines \\
(IREAD=1)}
\label{INPUTFUN1}
If IREAD=1 is chosen, an input distribution(s) at $Q_0^2$
should be supplied in the end of the FORTRAN program BFP1
as a function subroutine(s).

\vspace{0.3cm}
\noindent
1) Nonsinglet case

An initial  nonsinglet-quark distribution
at $Q_0^2$ should be given in QNS0(X)
as a double precision function.
As an example, the GS (set A) valence quark distribution
$x \Delta u_v+x \Delta d_v$ \cite{GS} at $Q_0^2$=4 GeV$^2$ is given
in the program BFP1.

\vspace{0.3cm}
\noindent
2) Singlet case

An initial singlet-quark distribution at $Q_0^2$ should be supplied in QS0(X),
and an initial gluon distribution in the nucleon should be in G0(X).
The GS $x\Delta q_{_S}=x\Delta u_v+x\Delta d_v+6x\Delta S$
and $x\Delta g$ distributions are given in the program.

\vspace{0.3cm}
\noindent
3) $\Delta q_i^+$ distribution case

In addition to the above $x\Delta q_{_S}$ and $x\Delta g$ distributions,
the initial $x\Delta q_i^+$ distribution 
(and another flavor distribution $x \Delta q_j^+$) should be 
supplied. In calculating two-flavor distributions simultaneously
for obtaining $x g_1$, two distributions (e.g. $x \Delta d^+$ and
$x \Delta s^+$) should be supplied in the functions QI0(1,x) and QI0(2,x).
If one needs only one-flavor evolution, one may set QI0(2,x)=0.
The GS distributions $x \Delta d^+$ and $x \Delta s^+$
are given in our program as an example.

%%%%%%%%%%%%%%%%%%%%%%%%%%%%%%%%%%%%%%%%%%%%%%%%%%%%%%%%%%%%%%%%%%%%%%%%%%%%%%
\subsection{Input distributions supplied by data files (IREAD=2)}
\label{INPUTFUN2}

If IREAD=2 is chosen, an input distribution(s) at $Q_0^2$
should be supplied in a separate data file(s).

\vspace{0.3cm}
\noindent
1) Nonsinglet case (data file \#13)

An initial  nonsinglet-quark distribution
at $Q_0^2$ should be given in the data file \#13 
as shown in the following example.

\begin{tabbing}
      0.000100 \ \ \ \ \ \ \= 0.009761 \kill
      0.000100   \>  0.009761 \\
      0.000110   \>  0.010184 \\
      0.000120   \>  0.010624 \\
      0.000132   \>  0.011081 \\
      0.000145   \>  0.011555 \\
       ...       \>   ...     \\
       ...       \>   ...     \\
       ...       \>   ...     \\
      1.000000   \>  0.000000 \\
\end{tabbing}

\vspace{-0.3cm}
The first column is the $x$ values and the second
one is the corresponding $x\Delta u_v+x\Delta d_v$ values.
The data at $x\le x_{min}$ and at $x$=1.0 must be supplied.

\vspace{0.3cm}
\noindent
2) Singlet case in the nucleon (data file \#14)

An initial singlet-quark distribution and a gluon distribution 
should be given in the data file \#14 as shown in the following.

\begin{tabbing}
      0.000100 \ \ \ \ \ \ \= 0.001721 \ \ \ \ \ \ \= 0.006729 \kill
      0.000100   \> 0.001721  \> 0.006729 \\
      0.000110   \> 0.001612  \> 0.007193 \\
      0.000120   \> 0.001485  \> 0.007688 \\
      0.000132   \> 0.001339  \> 0.008218 \\
      0.000145   \> 0.001172  \> 0.008784 \\
       ...       \>  ...      \>   ...    \\
       ...       \>  ...      \>   ...    \\
       ...       \>  ...      \>   ...    \\
      1.000000   \> 0.000000  \> 0.000000 \\
\end{tabbing}

\vspace{-0.3cm}
The first column is the $x$ values, the second is
the $x \Delta q_{_S}$ distribution, and the third is
the $x \Delta g$ distribution.
The data at $x\le x_{min}$ and at $x$=1.0 must be supplied.

\vspace{0.3cm}
\noindent
3) $\Delta q_i^+$ distribution case (data file \#15)

An initial $\Delta q_i^+$ distribution, a singlet-quark distribution,
and a gluon distribution should be given in the data file \#15
as shown in the following.

\begin{tabbing}

0.000100 \ \ \ \ \ \ \= -0.003637 \ \ \ \ \ \ \= -0.002681
\ \ \ \ \ \ \= 0.001721 \ \ \ \ \ \ \= 0.006729 \kill
      0.000100 \> -0.003637 \> -0.002681 \> 0.001721 \> 0.006729 \\
      0.000110 \> -0.003884 \> -0.002859 \> 0.001612 \> 0.007193 \\
      0.000120 \> -0.004147 \> -0.003048 \> 0.001485 \> 0.007688 \\
      0.000132 \> -0.004428 \> -0.003249 \> 0.001339 \> 0.008218 \\
      0.000145 \> -0.004727 \> -0.003462 \> 0.001172 \> 0.008784 \\
      ...      \>  ...      \>  ...      \>   ...    \>  ...     \\
      ...      \>  ...      \>  ...      \>   ...    \>  ...     \\
      ...      \>  ...      \>  ...      \>   ...    \>  ...     \\
     1.000000  \> 0.000000  \> 0.000000  \> 0.000000 \> 0.000000 \\
\end{tabbing}

\vspace{-0.3cm}
The first column is the $x$ values,
the second is the $x \Delta q_i^+$ distribution,
the third is another flavor distribution $x \Delta q_j^+$,
the fourth is $x \Delta q_{_S}$, and
the fifth is $x \Delta g$.
The data at $x\le x_{min}$ and at $x$=1.0 must be supplied.
If one needs only one-flavor evolution, the second or third column
values are set to 0.

\vfill\eject
%%%%%%%%%%%%%%%%%%%%%%%%%%%%%%%%%%%%%%%%%%%%%%%%%%%%%%%%%%%%%%%%%%%%%%%%%%%%%%
%%%%%%%%%%%%%%%%%%%%%%%%%%%%%%%%%%%%%%%%%%%%%%%%%%%%%%%%%%%%%%%%%%%%%%%%%%%%%%
\section{{\bf Description of the program BFP1}}\label{BFP1}
\setcounter{equation}{0}
\setcounter{figure}{0}
\setcounter{table}{0}

The major part of the program BFP1 is essentially the same as
the unpolarized version \cite{BF}. We do not explain each subroutine
in this section. They are already well described in the previous
publication, so that the interested reader may read Ref. \cite{BF}
for the detailed description of all the subroutines.

The main program reads fifteen input parameters 
from the input file \#10.
Actual $Q^2$ evolution calculations are done by calling
the subroutine GETQNS in the nonsinglet case
and GETQS in the singlet (or $x\Delta q_i$) case.
Each evolution step is calculated by the subroutine QNSXT or GETQGX.
Minor modifications are made in each flavor evolution (GETQS)
so that two quark distributions ($x\Delta q_i^+$ and $x\Delta q_j^+$)
are evolved simultaneously.
A typical function of the GS parton distributions is
given as the function GS(X,A,B,C,D,E,F).
We explain the subroutines associated with the
input distributions in the following.

\vspace{0.4cm}

\subsection{Function GS(X,A,B,C,D,E,F)}  
\vspace{0.3cm}

This function calculates the GS parton distributions given
by the parameters A, B, C, D, E, and F as
$ABx^C(1-x)^D(1+Ex+F\sqrt{x})$.

\vspace{0.4cm}

\subsection{Functions QNS0(X), QS0(X), G0(X), and QI0(I,X)}   
\vspace{0.3cm}

The functions QNS0, QS0, and G0 calculate 
an initial nonsinglet-quark distribution $x\Delta q_{_{NS}}$,
a singlet-quark distribution $x\Delta q_{_S}$, 
and a gluon distribution $x\Delta g$. 
As an example, the GS distributions \cite{GS} are provided. \\ 
The nonsinglet one is 
\begin{align}
QNS0  =& x\Delta u_v+x\Delta d_v  \nonumber \\
      =& 0.918*1.365 x^{0.512} (1-x)^{3.96} (1+11.65x-4.6\sqrt{x}) \nonumber \\
       &+ (-0.339)*3.849 x^{0.78} (1-x)^{4.96} (1+7.81x-3.48\sqrt{x}) 
\ ,
\end{align}
the singlet one in the three flavor case is
\begin{align}
QS0  =& x\Delta u_v+x\Delta d_v+ 6 x\Delta S \nonumber \\
     =& 0.918*1.365 x^{0.512} (1-x)^{3.96} (1+11.65x-4.6\sqrt{x}) \nonumber \\
      &+ (-0.339)*3.849 x^{0.78} (1-x)^{4.96} (1+7.81x-3.48\sqrt{x}) \nonumber \\
      &+ 6*(-0.06)*18.521 x^{0.724} (1-x)^{14.4} (1+4.63x-4.96\sqrt{x}) 
\ ,
\end{align}
and the gluon distribution is
\beqn
G0 = 1.71*3.099 x^{0.724} (1-x)^{5.71} (1+0.0x+0.0\sqrt{x}) 
\ .
\eeqn

The function QI0 calculates two initial flavor distributions.
As an example, the GS 
$x\Delta d^+ = x\Delta d +x\Delta \overline d$ and 
$x\Delta s^+ = x\Delta s +x\Delta \overline s$ distributions
are provided:
\begin{align}
QI0_1 =& x\Delta d_v + 2 x \Delta \bar d \nonumber \\
      =& -0.339*3.849 x^{0.78} (1-x)^{4.96} (1+7.81x-3.48\sqrt{x}) \nonumber \\
      &+ 2*(-0.06)*18.521 x^{0.724} (1-x)^{14.4} (1+4.63x-4.96\sqrt{x}) 
\ ,
\end{align}
\begin{align}
QI0_2 =& 2x\Delta s \nonumber \\
      =& 2*(-0.06)*18.521 x^{0.724} (1-x)^{14.4} (1+4.63x-4.96\sqrt{x})
\ .
\end{align}
In these equations, the flavor symmetry 
$\Delta \bar u=\Delta \bar d=\Delta \bar s=\Delta S$ is assumed.

\vfill\eject
%%%%%%%%%%%%%%%%%%%%%%%%%%%%%%%%%%%%%%%%%%%%%%%%%%%%%%%%%%%%%%%%%%%%%%%%%%%%%%
%%%%%%%%%%%%%%%%%%%%%%%%%%%%%%%%%%%%%%%%%%%%%%%%%%%%%%%%%%%%%%%%%%%%%%%%%%%%%%
\section{{\bf Numerical analysis}}\label{RESULTS}
\setcounter{equation}{0}
\setcounter{figure}{0}
\setcounter{table}{0}

%%%%%%%%%%%%%%%%%%%%%%%%%%%%%%%%%%%%%%%%%%%%%%%%%%%%%%%%%%%%%%%%%%%%%%%%%%%%%%
\subsection{Accuracy of $Q^2$ evolution results}\label{ACCURA}

There are two parameters which determine the accuracy of our numerical
results. They are the numbers of steps: $N_x$ and $N_t$.
As they become larger, the accuracy becomes better.
However, there is a restriction because
the evolution computation should be done within a certain
CPU time depending on the machine power.
We check how the evolution results depend on these parameters
and how long it takes for finishing the evolution.

First, we fix the $x$ step at $N_x$=1000, then the $t$ step $N_t$
is varied as 20, 50, 200, and 1000. Numerical results for the 
polarized singlet quark distribution are shown in Fig. \ref{fig:dqs-nt}.
The initial distribution is the GS-A at $Q^2$=4 GeV$^2$
\cite{GS}. The three-flavor distribution is considered.
Because there is little information on each antiquark distribution \cite{SK},
flavor symmetric distributions are assumed for the antiquark distributions.
The scale parameter is taken as $\Lambda$=231 MeV in fitting
polarized experimental data.
We use $N_f=$4 and $\Lambda$=231 MeV for calculating the evolution.
The GS-A singlet distribution is evolved to the one at $Q^2$=200 GeV$^2$ 
by our NLO program. The input distribution is supplied in the end of
the program as the subroutine QS0.
The input parameters are 
IOUT=3, IREAD=1, INDIST=1, IORDER=2, ITYPE=2, IMORP=2,
Q02=4.0, Q2=200.0, DLAM=0.231, NF=4, XX=0.0,
NX=1000, NT=200, NSTEP=100, and NXMIN=$-$4.
There is almost no difference between the various results 
in Fig. \ref{fig:dqs-nt}, which
means that merely fifty steps are enough for getting accurate evolution.
This is certainly expected from the fact that the scaling violation
is a small logarithmic effect. 

Next, the step $N_x$ is varied with fixed $N_t$=200 in Fig. \ref{fig:dqs-nx}.
$N_x$=100, 300, 1000, and 4000 are taken. Even in the 300 steps,
we obtain rather accurate results. 
The evolution results become slightly worse in the gluon case
as shown in Figs. \ref{fig:dg-nt} and \ref{fig:dg-nx}.
Defining the evolution accuracy by 
$| \, [ \Delta p (N_x,N_t) - \Delta p (N_x=4000,N_t=1000) ]
 / \Delta p (N_x=4000,N_t=1000) \, |$,
we find that the accuracy is better than 1\% 
in the region $10^{-5}<x<0.8$ with $N_x=1000$ and $N_t=200$.
However, it takes rather a significant amount of running CPU time.
It is typically seven minutes on the AlphaServer 2100 4/200
in a nonsinglet case and sixty minutes in a singlet case.
Our program is good enough for a single evolution calculation, but it
is not efficient enough for repeated use. We will work on
a much faster program in future.

We also check our results with other publications.
First, the evolution in Ref. \cite{GS} from $Q^2$=4 GeV$^2$
to $Q^2$=50 GeV$^2$ are compared with our results.
The structure functions $x g_1^p$ and $x g_1^n$ at $Q^2$=50 GeV$^2$
agree well with our evolved ones.
Second, the evolution in Ref. \cite{WM} is compared with ours.
The GS-A input distributions are used. 
They are evolved to the ones at $Q^2$=10 GeV$^2$ with $\Lambda$=239 MeV
and are also devolved to the ones at $Q^2$=2 GeV$^2$.
The singlet distribution $\Delta q_s$ and the gluon distribution 
$\Delta g$ are compared. We find that they agree well with 
our evolution results. From these studies on our results
in comparison with others and from repeated checks on our
evolution program, we find that the program BFP1 is reliable.
We should mention that there is another work \cite{OTHERG1}
on the NLO evolution in addition to Refs. \cite{GS} and \cite{WM}.

%%%%%%%%%%%%%%%%%%%%%%%%%%%%%%%%%%%%%%%%%%%%%%%%%%%%%%%%%%%%%%%%%%%%%%%%%%%%%%
\subsection{Calculation of $g_1$ from output data}\label{makeg1}

We do not supply the BFP1 program so that the structure function
$g_1$ is obtained directly in the output file. It is because $g_1$ depends
on the number of flavor and because the precise evolution should be
described by setting flavor thresholds. We ask the reader to set up
these points by oneself with our program.
Actual evolution results for $g_1$ are shown in the next subsection,
where the following prescription is used for calculating $g_1$
from the output data file \#11.

First, IOUT=5 or 6 should be chosen so that 
the evolution of $\Delta q_{i}^+(x)$, $\Delta q_{j}^+(x)$,
$\Delta q_{_S}(x)$, and $\Delta g(x)$ is calculated
by Eqs. (\ref{singlet}) and (\ref{flavorap}).
Furthermore, ITYPE=1 should be chosen for calculating
the convolution integrals of these evolved distributions with
the coefficient functions in Eq. (\ref{g1nlo}). 
The convolution results are written in the output file \#11,
and we denote them as
$g_{1, d}^+(x)$, $g_{1, str}^+(x)$, $g_{1_S}^+(x)$, and
$g_{1, g}^+(x)$.
It goes without saying that the convolution is not necessary in the LO case.
The LO $g_1$ is calculated directly by Eq. (\ref{g1lo}).
Next, the $g_1$ for the proton is, for example, calculated by choosing
$\Delta d^+ (x)$ and $\Delta s^+ (x)$ for the distributions
$\Delta q_{i}^+(x)$ and $\Delta q_{j}^+(x)$:
\begin{align}
g_1^p (x,Q^2) &= \frac{1}{2} \, \sum_i^{N_f} e_i^2 \, 
                   \left [ \, g_{1, i} (x,Q^2) 
                           +  g_{1, g} (x,Q^2) \, \right ]
\nonumber \\
&= \frac{1}{18} \, \left [ \, 4 \, g_{1_S}^+(x,Q^2) 
                       - 3 \, \{ g_{1, d}^+ (x,Q^2) + g_{1, str}^+ (x,Q^2) \}
                     \, \right ]
             + \frac{1}{2} \, \sum_i^{N_f} e_i^2 \, g_{1, g} (x,Q^2) 
\ ,
\end{align}
in the three and four flavor cases. Five and six flavor evolution can
be calculated in the similar way with a slightly modified equation.

%%%%%%%%%%%%%%%%%%%%%%%%%%%%%%%%%%%%%%%%%%%%%%%%%%%%%%%%%%%%%%%%%%%%%%%%%%%%%%
\subsection{Comparisons with experimental data}\label{COMP}

We calculate evolution of the structure function $g_1$ and
compare its results with experimental data.
Using the procedure in the last subsection,
we obtain the evolution results for $g_1$.
The LO and NLO results are shown in Fig. \ref{fig:dg1}.
The same GS-A polarized parton distributions are used as
the input distributions at 
$Q^2$=4 GeV$^2$ in both LO and NLO cases. To be precise, 
this is not a correct procedure because the GS-A is set up
for the NLO $\overline{MS}$ calculation. It should not be used
in the LO calculation. Nevertheless, the GS-A distributions are
also employed in the LO for finding
NLO effects on $g_1$. Therefore, the differences between the
solid and dashed curves at $Q^2$=4 GeV$^2$ are purely
due to the coefficient functions.
Because the coefficient-function contributions are rather large,
it is difficult to find the NLO evolution effects from Fig. \ref{fig:dg1}.
In order to illustrate the NLO effects from $Q^2$=4 GeV$^2$ to
$Q^2$=200 GeV$^2$, we show the evolution of the distribution
$\sum_i e_i^2 \, x \, \Delta q_i /2$ in Fig. \ref{fig:dqg1}.
The same distributions are taken at $Q^2$=4 GeV$^2$
in the LO and NLO cases.
The figure indicates that the NLO effects are positive 
in the small and large $x$ regions and they are negative
in the intermediate region $x\approx 0.2$.

Our evolution results are compared with spin-asymmetry data
in Figs. \ref{fig:a-1}, \ref{fig:a-2}, and \ref{fig:a-3}.
The SLAC-E130 \cite{E130}, SLAC-E143 \cite{E143}, EMC \cite{EMC}, 
and SMC \cite{SMC} data are shown in the figures.
We comment how the spin asymmetry is calculated theoretically.
The asymmetry is given by the structure functions $g_1$ and $F_1$ as
\beqn
A_1 \cong \frac{g_1(x,Q^2)}{F_1(x,Q^2)} 
  = g_1(x,Q^2) \, \frac{2\, x\, (1+R)}{F_2(x,Q^2)} \ .
\label{geta}
\eeqn
The function $R$ is given by $R=(F_2-2xF_1)/(2xF_1)$,
and it could be taken from Ref. \cite{R1990}.
In fitting experimental data for obtaining the optimum
unpolarized parton distributions, the $F_2$ structure function
is used instead of $F_1$. Therefore, it is 
better to calculate the asymmetry with 
$F_2$ if we would like to compare with the experimental data.
The MRS-G unpolarized parton distributions \cite{MRSG} are used
for calculating $F_2$ and the function R in Ref. \cite{R1990}
is used in Eq. (\ref{geta}).

In Fig. \ref{fig:a-1}, our evolution curves at $x$=0.035 are shown
with the asymmetry $A_1$ data. 
The dashed and solid curves indicate the LO and NLO evolution 
results respectively. In the large $Q^2$ region, both results are almost
the same; however, they differ significantly at small $Q^2$ in particular
in the region $Q^2<\, $2 GeV$^2$.
The difference is not so large at slightly larger $x$ (=0.08) as shown
in Fig. \ref{fig:a-2}. However, the LO values become larger than those of
the NLO evolution. In the medium $x$ region ($x$=0.25), the difference
becomes larger again at small $Q^2$ as shown in in Fig. \ref{fig:a-3}.
From these figures, we find that the asymmetry has $Q^2$ dependence
although it is not large. People used to assume that the asymmetry
is independent of $Q^2$ by neglecting the $Q^2$ evolution difference
between $g_1$ and $F_1$ in analyzing the experimental data.
We find clearly that it is not the case. For a precise analysis,
the $Q^2$ dependence in the asymmetry has to be taken into account.
Currently, parametrization studies are in progress \cite{RHIC-J}
by using our program. We expect to obtain NLO fitting results 
in the near future.

\vfill\eject
%%%%%%%%%%%%%%%%%%%%%%%%%%%%%%%%%%%%%%%%%%%%%%%%%%%%%%%%%%%%%%%%%%%%%%%%%%%%%%
%%%%%%%%%%%%%%%%%%%%%%%%%%%%%%%%%%%%%%%%%%%%%%%%%%%%%%%%%%%%%%%%%%%%%%%%%%%%%%
\section{{\bf Summary}}\label{SUMMARY}
\setcounter{equation}{0}
\setcounter{figure}{0}
\setcounter{table}{0}

We have investigated numerical solution of the spin-dependent DGLAP
equations with or without the NLO corrections. The solution method is 
so called brute-force method. A FORTRAN program is provided for evolving
longitudinally polarized parton distributions including nonsinglet, singlet,
each flavor, and gluon distributions. Furthermore, the evolution results
could be written in a structure-function form, which is
the convolution of the evolved distributions with the coefficient functions. 
Therefore, the structure function $g_1$ could be also calculated from our
output data.
We checked that typical accuracy is better than 1\% with reasonable
numbers of steps: $N_x$=1000 and $N_t$=200. Comparisons with experimental
data indicate significant $Q^2$ dependence and NLO effects in the region
$Q^2\approx 1$ GeV$^2$. Therefore, the $Q^2$ independence assumption,
which was used to be employed in analyzing $A_1$ experimental data,
is not valid. We have to include the $Q^2$ evolution differences
between $F_1$ and $g_1$. Our evolution code is very useful for theoretical
and experimental researchers in high-energy spin physics.

\vspace{0.7cm}
%%%%%%%%%%%%%%%%%%%%%%%%%%%%%%%%%%%%%%%%%%%%%%%%%%%%%%%%%%%%%%%%%%%%%%%%%%%%%%
%%%%%%%%%%%%%%%%%%%%%%%%%%%%%%%%%%%%%%%%%%%%%%%%%%%%%%%%%%%%%%%%%%%%%%%%%%%%%%
\section*{{\bf Acknowledgments}}
\addcontentsline{toc}{section}{\protect\numberline{\S}{Acknowledgments}}

MH, SK, and MM thank the Research Center for Nuclear Physics 
in Osaka for making them use computer facilities.
They thank participants, in particular those in Ref. \cite{RHIC-J},
of the RHIC-SPIN-J collaboration meetings
for discussions on the polarized parton distributions.

\vfill\eject
%%%%%%%%%%%%%%%%%%%%%%%%%%%%%%%%%%%%%%%%%%%%%%%%%%%%%%%%%%%%%%%%%%%%%%%%%%%%%%
%%%%%%%%%%%%%%%%%%%%%%%%%%%%%%%%%%%%%%%%%%%%%%%%%%%%%%%%%%%%%%%%%%%%%%%%%%%%%%
\section*{{\bf Appendix A. Splitting functions}}
\addcontentsline{toc}{section}{\protect\numberline{\S}
{Appendix A. Splitting functions}}
\renewcommand{\theequation}{A.\arabic{equation}}

First, the constants $\beta_0$, $\beta_1$ 
$C_G$, $C_F$, and $T_R$ are given by
the number of color ($N_c$=3) and the number of flavor ($N_f$) as
\beqn
\beta_0 = \frac{11}{3} C_G-\frac{4}{3}T_R N_f  
\ , \ \ \ \ 
\beta_1 = \frac{34}{3} C^2_G - \frac{10}{3} C_G N_f - 2 C_F N_f  
\ ,
\eeqn
\beqn 
C_G=N_c
\ , \ \ \  C_F=\frac{N_c^2-1}{2N_c}
\ , \ \ \ T_R=\frac{1}{2}
\ .
\eeqn

Splitting function in the leading order(LO) are
\begin{align}
\Delta P_{NS}^{(0)}(x) &= C_F \, \left [ \, \frac{2}{(1-x)_+}-1-x 
+\frac{3}{2} \, \delta(1-x) \, \right ] 
\ ,
\nonumber \\
\Delta P_{q_i^+ q_j^+}^{(0)}(x) &= \delta_{ij} \, \Delta P_{NS}^{(0)}(x)
\ ,
\nonumber \\
\Delta P_{q_i^+ g}^{(0)}(x) &= 2 \, T_R \, (2x-1)  
\ ,
\nonumber \\
\Delta P_{g q_i^+}^{(0)}(x) &= C_F \, (2-x)  
\ , 
\nonumber \\
\Delta P_{gg}^{(0)}(x) &= 2 \, C_G \, 
      \left[ \, \frac{1}{(1-x)_+}-2 x+1 \, \right]
        +\frac{\beta_0}{2} \, \delta(1-x) 
\ ,
\end{align}
where the $+$ function is defined by
\beqn
\int_0^1 dx \, {\frac{f(x)}{(1-x)_+}} \ = \ 
\int_0^1 dx \, {\frac{f(x)-f(1)}{1-x}}
\ .
\label{pfun}
\eeqn
It should be noted that the above integration is defined
in the region $0\le x\le 1$.

The NLO splitting functions for $\Delta q^+_i(x)$ and $\Delta g(x)$ 
are given by \cite{PNLO}
\begin{align}
\Delta P^{(1)}_{q^+_{i} q^+_j} (x) &= \delta_{ij} \, \Delta P^{(1)}_{q^+,NS} 
                                  + 2 \, C_F \, T_R \, F_{qq} 
\ , \nonumber \\
\Delta P^{(1)}_{q^+_{i} g} (x) &=
       C_F \, T_R \, F^1_{qg}(x) + C_G \, T_R \, F^2_{qg}(x) 
\ , \nonumber \\
\Delta P^{(1)}_{g q^+_i} (x) &= 
       C_F \, T_R \, N_f \, F^1_{gq}(x) 
      +C^2_F \, F^2_{gq}(x) + C_F \, C_G \, F^3_{gq}(x)  
\ , \\
\Delta P^{(1)}_{g g} (x) &= 
      - C_G \, T_R \, N_f \, F^1_{gg}(x) 
      - C_F \, T_R \, N_f \, F^2_{gg}(x) + C^2_G \, F^3_{gg}(x)
\ . \nonumber 
\end{align}
The splitting functions in the singlet equations (\ref{singlet})
are expressed as
\begin{align}
\Delta P_{qq} \otimes \Delta q_s  &= 
        \sum_{i,j} \Delta P_{q^+_{i} q^+_j} \otimes \Delta q_j \ , \ \
\Delta P_{qg} \otimes \Delta g   = 
        \sum_{i}   \Delta P_{q^+_{i} g} \otimes \Delta g \ , 
\nonumber \\
\Delta P_{gq} \otimes \Delta q_s &= 
        \sum_{j}   \Delta P_{g q^+_j} \otimes \Delta q_j \ .
\end{align}

The splitting function $\Delta P^{(1)}_{q^\pm,NS}$ is given
\beqn
 \Delta P^{(1)}_{q^\pm,NS} = P^{(1)}_{q^\mp,NS} \ ,
\eeqn
where $P^{(1)}_{q^\mp,NS}$ 
are the unpolarized NLO nonsinglet splitting functions.
The $\pm$ in these equations indicates
``$\Delta q \pm \Delta \bar q$ type" distribution 
$\sum_i a_i (\Delta q_i \pm \Delta \bar q_i)$.
The function $\Delta P^{(1)}_{q^\pm,NS}$ is given by
\begin{align}
\Delta P^{(1)}_{q^\pm,NS} (x) 
    =&C_F^2\biggl\{P_F(x) \mp P_A(x)+\delta(1-x)
    \biggl[\frac{3}{8}-\frac{1}{2}\pi^2
    +\zeta(3)-8\widetilde{S}(\infty)\biggr]\biggr\}  
\nonumber \\
&+\frac{1}{2}C_FC_A\bigg\{P_G(x) \pm P_A(x)+\delta(1-x)
    \biggl[\frac{17}{12}+\frac{11}{9}
    \pi^2-\zeta(3)+8\widetilde{S}(\infty)\biggr]\biggr\}  
\nonumber \\
&+C_FT_RN_f\biggl\{P_{N_F}(x)-\delta(1-x)
    \biggl(\frac{1}{6}+\frac{2}{9}\pi^2\biggr)
    \biggr\}  
\ ,
\end{align}
where $P_F(x)$, $P_G(x)$, $P_{N_F}(x)$, and $P_A(x)$ are given
in Ref. \cite{CFP}
\begin{align}
P_F(x) &= -2{ \frac{1+x^2}{1-x} }\ln x\ln(1-x)-
\biggl({\frac{3}{1-x}}+2x\biggr)\ln x- \frac{1}{2} (1+x)\ln^2x  -\!\!5(1-x) 
\ , \nonumber \\
P_G(x) &= \frac{1+x^2}{(1-x)_+} \bigg[\ln^2x+\frac{11}{3}\ln x+
\frac{67}{9}
-\frac{1}{3}\pi^2\biggr]+2(1+x)\ln x+\frac{40}{3}(1-x)  
\ , \nonumber \\
P_{N_F}(x) &= \frac{2}{3}\Biggl[ \frac{1+x^2}{(1-x)_+}\biggl(
-\ln x-\frac{5}{3}\biggr)-2(1-x)\Biggr] 
\ , \nonumber \\
P_A(x)  &= 2 \frac{1+x^2}{1+x}\int_{x/(1+x)}^{1/(1+x)} \frac{dz}{z}
\ln \frac{1-z}{z}
+2(1+x)\ln x+4(1-x) 
\ .
\end{align}
The functions $F_{qq}$, $F_{qg}$, $F_{gq}$, and $F_{gg}$ are defined by
\begin{equation}
F_{qq} (x) = (1-x)-(1-3 x) \ln x-(1 + x) \ln^2 x
\ , 
\end{equation}
\begin{align}
F^1_{qg} (x) =& -22 + 27x - 9 \ln x + 8(1 - x) \ln (1-x) \nonumber \\
             &+ \delta p_{qg}(x) \left[2 \ln^2 (1-x)
                 - 4 \ln (1-x) \ln x + \ln^2 x - \frac{2}{3} \pi^2 \right] 
\ ,\nonumber \\
F^2_{qg} (x) =& \, 2 (12 - 11x) - 8(1 - x) \ln (1-x) + 2(1 + 8x) \ln x 
\nonumber \\
             &- 2 \left[\ln^2 (1-x) - \frac{\pi^2}{6} \right] \delta p_{qg}(x) 
                 - \left[2 S_2 (x) - 3 \ln^2 x \right] \delta p_{qg}(-x) 
\ ,
\end{align}
\begin{align}
F^1_{gq} (x) =& -\frac{4}{9}(x+4)-\frac{4}{3} \delta p_{gq}(x) \ln (1-x) 
\ ,\nonumber \\
F^2_{gq} (x) =& -\frac{1}{2}-\frac{1}{2}(4-x)\ln x 
                 -\delta p_{gq}(-x)\ln (1-x) 
              + \left[-4 - \ln^2(1-x) + \frac{1}{2} \ln^2 x \right] 
                 \delta p_{gq}(x) 
\ ,\nonumber \\
F^3_{gq} (x) =& \, (4-13x) \ln x + \frac{1}{3}(10 + x) \ln (1-x) 
                 + \frac{1}{9}(41 + 35x)
\nonumber \\
             & + \frac{1}{2}[-2 S_2 (x) + 3 \ln^2 x] \delta p_{gq}(-x) 
\nonumber \\
             &+ \left[\ln^2 (1-x) - 2 \ln (1-x) \ln x 
                - \frac{\pi^2}{6} \right] \delta p_{gq}(x) 
\ ,
\end{align}
\begin{align}
F^1_{gg} (x) =& \, 4(1 - x)+\frac{4}{3}(1 + x) \ln x 
                 + \frac{20}{9} \delta p_{gg}(x)+\frac{4}{3} \delta (1-x) 
\ ,\nonumber \\
F^2_{gg} (x) =& \, 10(1-x) + 2(5-x) \ln x + 2(1+x) \ln^2 x + \delta (1-x) 
\ ,\nonumber \\
F^3_{gg} (x) =& \, \frac{1}{3}(29-67x) \ln x -\frac{19}{2}(1-x) + 4(1+x) 
                  \ln^2 x -2 S_2 (x) \delta p_{gg}(-x) 
\nonumber \\
             &+ \left[\frac{67}{9}-4\ln (1-x)\ln x + \ln^2 x - \frac{\pi^2}{3}              
    \right] \delta p_{gg}(x) + \left[3 \zeta(3) 
                 +\frac{8}{3} \right] \delta (1-x) 
\ , 
\end{align}
where,  $\delta p_{qg}\ , \delta p_{gq}\ , \delta p_{gg}$ are defined 
\begin{align}
\delta p_{qg} (x) &\equiv 2 x-1 \ ,\nonumber \\
\delta p_{gq} (x) &\equiv 2 - x \ ,\\
\delta p_{gg} (x) &\equiv \frac{1}{(1-x)_+} - 2 x + 1 \ . \nonumber
\end{align}
The $\zeta$ function is defined by
$\zeta (k)=\sum_{n=1}^\infty \frac{1}{n^k}$,
and $\zeta(3)$ is given by the numerical value ($\zeta(3)$=1.2020569...). 
The $S_2$ function is expressed in terms of the Spence function $S(x)$
\begin{align}
S_2(x) &= \int_{x/(1+x)}^{1/(1+x)} \frac{dz}{z} \ln {\frac{1-z}{z}} 
\nonumber \\
       &= S\left({\frac{x}{1+x}}\right) - S\left({\frac{1}{1+x}}\right) 
           - \frac{1}{2} \left[\ln^2 \frac{1}{1+x}
           - \ln^2 \frac{x}{1+x} \right] 
\ ,
\end{align}
where $S(x)$ is defined by 
\beqn
S(x) \, = \, \int_x^1 dz \, \frac{\ln z}{1-z} \ .
\eeqn
It should be noted that another convention is sometimes used,
namely $-S(x)$ may be called the Spence function.
It is useful to use a series expansion form for numerical
calculations
\beqn
S(x) \, = \, - \, \sum_{k=1}^\infty \frac{(1-x)^k}{k^2}
\ .
\eeqn

%\vspace{0.7cm}
\vfill\eject
%%%%%%%%%%%%%%%%%%%%%%%%%%%%%%%%%%%%%%%%%%%%%%%%%%%%%%%%%%%%%%%%%%%%%%%%%%%%%%
%%%%%%%%%%%%%%%%%%%%%%%%%%%%%%%%%%%%%%%%%%%%%%%%%%%%%%%%%%%%%%%%%%%%%%%%%%%%%%
\section*{{\bf Appendix B. Coefficient functions}}
\addcontentsline{toc}{section}{\protect\numberline{\S}
{Appendix B. Coefficient functions}}
\renewcommand{\theequation}{B.\arabic{equation}}

The coefficient functions $\Delta C_{q}$ and $\Delta C_{q}$
are \cite{PNLO}

\begin{align}
\Delta C_{q}(x,\alpha_s) &= \delta(1-x)  + 
                \frac{\alpha_s}{2\pi} \, \Delta B_{q}(x)
\ , \\ 
\Delta C_{g}(x,\alpha_s) &= \frac{\alpha_s}{2\pi} \, \Delta B_{g}(x)
\ ,
\end{align}
where the functions $\Delta B_q$ and $\Delta B_g$ are

\begin{align}
\Delta B_{q}(x) =& \, 
C_F \Bigg[(1+x^2) \left\{ \frac{\ln (1-x)}{1-x} \right\}_+
                          - \frac{3}{2} \frac{1}{(1-x)_+}
                          - \frac{1+x^2}{1-x}\ln x
\nonumber \\
&+ 2+x-\left(\frac{9}{2}+\frac{\pi^2}{3}  \right) \delta (1-x) \Bigg]
\ , 
\end{align}
\begin{align}
\Delta B_{g}(x) =& \, 
2 \, T_R \left[(2x-1) \left(\ln \frac{1-x}{x} -1 \right) + 2(1-x) \right]
\ .
\end{align}
\vfill\eject
%%%%%%%%%%%%%%%%%%%%%%%%%%%%%%%%%%%%%%%%%%%%%%%%%%%%%%%%%%%%%%%%%%%%%%%%%%%%%%
%%%%%%%%%%%%%%%%%%%%%%%%%%%%%%%%%%%%%%%%%%%%%%%%%%%%%%%%%%%%%%%%%%%%%%%%%%%%%%

\vfill\eject
%%%%%%%%%%%%%%%%%%%%%%%%%%%%%%%%%%%%%%%%%%%%%%%%%%%%%%%%%%%%%%%%%%%%%%%%%%%%%%
%%%%%%%%%%%%%%%%%%%%%%%%%%%%%%%%%%%%%%%%%%%%%%%%%%%%%%%%%%%%%%%%%%%%%%%%%%%%%%
\section*{{\bf TEST RUN OUTPUT}}
\addcontentsline{toc}{section}{\protect\numberline{\S}
{TEST RUN OUTPUT}}

\noindent
IOUT= 3\ \ \ IREAD= 1\ \ \ INDIST= 1 IORDER= 2\ \ \ 
ITYPE= 2\ \ \ IMORP= 2 \\
Q02= 4.0000\ \ \ Q2= 200.000\ \ \ DLAM= 0.2310\ \ \ NF= 4 \\
XX= 0.0000000\ \ \ NX=1000\ \ \ NT= 200\ \ \ NSTEP=  50\ \ \ NXMIN= $-$4 \\

\begin{tabbing}
     0.000100 \ \ \ \ \= -0.019166 \ \ \ \ \= 0.097380 \=  \kill
     0.000100 \>  -0.019166  \>  0.097380 \> \\
     0.000120 \>  -0.020672  \>  0.105178 \> \\
     0.000145 \>  -0.022276  \>  0.113534 \> \\
     0.000174 \>  -0.023982  \>  0.122478 \> \\
     0.000209 \>  -0.025790  \>  0.132045 \> \\
     0.000251 \>  -0.027703  \>  0.142266 \> \\
     0.000302 \>  -0.029718  \>  0.153174 \> \\
     0.000363 \>  -0.031833  \>  0.164803 \> \\
     0.000437 \>  -0.034043  \>  0.177185 \> \\
     0.000525 \>  -0.036338  \>  0.190351 \> \\
     0.000631 \>  -0.038707  \>  0.204330 \> \\
     0.000759 \>  -0.041130  \>  0.219150 \> \\
     0.000912 \>  -0.043584  \>  0.234833 \> \\
     0.001096 \>  -0.046036  \>  0.251398 \> \\
     0.001318 \>  -0.048444  \>  0.268856 \> \\
     0.001585 \>  -0.050754  \>  0.287209 \> \\
     0.001905 \>  -0.052898  \>  0.306451 \> \\
     0.002291 \>  -0.054790  \>  0.326559 \> \\
     0.002754 \>  -0.056324  \>  0.347493 \> \\
     0.003311 \>  -0.057372  \>  0.369192 \> \\
     0.003981 \>  -0.057777  \>  0.391566 \> \\
     0.004786 \>  -0.057353  \>  0.414496 \> \\
     0.005754 \>  -0.055883  \>  0.437819 \> \\
     0.006918 \>  -0.053115  \>  0.461325 \> \\
     0.008318 \>  -0.048770  \>  0.484750 \> \\
     0.010000 \>  -0.042545  \>  0.507761 \> \\
     0.012023 \>  -0.034126  \>  0.529949 \> \\
     0.014454 \>  -0.023214  \>  0.550821 \> \\
     0.017378 \>  -0.009552  \>  0.569787 \> \\
     0.020893 \>   0.007027  \>  0.586160 \> \\
     0.025119 \>   0.026539  \>  0.599147 \> \\
     0.030200 \>   0.048784  \>  0.607861 \> \\
     0.036308 \>   0.073276  \>  0.611331 \> \\
     0.043652 \>   0.099181  \>  0.608536 \> \\
     0.052481 \>   0.125309  \>  0.598449 \> \\
     0.063096 \>   0.150159  \>  0.580109 \> \\
     0.075858 \>   0.172073  \>  0.552726 \> \\
     0.091201 \>   0.189500  \>  0.515809 \> \\
     0.109648 \>   0.201326  \>  0.469336 \> \\
     0.131826 \>   0.207180  \>  0.413952 \> \\
     0.158489 \>   0.207504  \>  0.351177 \> \\
     0.190546 \>   0.203229  \>  0.283585 \> \\
     0.229087 \>   0.194990  \>  0.214868 \> \\
     0.275423 \>   0.182213  \>  0.149663 \> \\
     0.331131 \>   0.162801  \>  0.092992 \> \\
     0.398107 \>   0.134220  \>  0.049196 \> \\
     0.478630 \>   0.096195  \>  0.020506 \> \\
     0.575440 \>   0.053887  \>  0.005833 \> \\
     0.691831 \>   0.018721  \>  0.000825 \> \\
     0.831764 \>   0.001943  \>  0.000022 \> \\
     1.000000 \>   0.000000  \>  0.000000 \> \\
\end{tabbing}

\vfill\eject
%%%%%%%%%%%%%%%%%%%%%%%%%%%%%%%%%%%%%%%%%%%%%%%%%%%%%%%%%%%%%%%%%%%%%%%%%%%%%%
%%%%%%%%%%%%%%%%%%%%%%%%%%%%%%%%%%%%%%%%%%%%%%%%%%%%%%%%%%%%%%%%%%%%%%%%%%%%%%
\section*{{\bf Figures}}
\addcontentsline{toc}{section}{\protect\numberline{\S}{Figures}}

\vspace{2.0cm}
%%%%%%%%%%%%%%%%%%%%%%%%%%%%%%%% figure %%%%%%%%%%%%%%%%%%%%%%%%%%%%%%%%%%%%%%
\begin{figure}[h]
   \begin{center}
      \epsfig{file=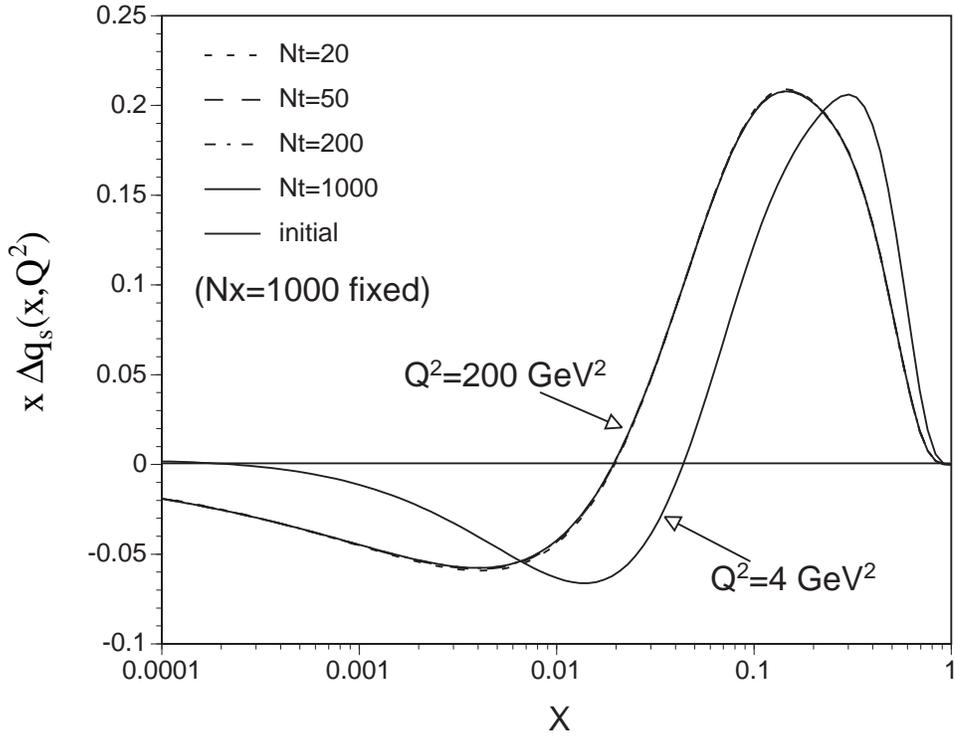,width=15.5cm}
   \end{center}
\caption{$N_t$ dependence of singlet evolution results is shown.
The initial distribution is the GS-A at Q$^2$=4 GeV$^2$.
It is evolved to the one at Q$^2$=200 GeV$^2$ by
the next-to-leading-order DGLAP evolution equations.
$N_x$=1000 is fixed and $N_t$ is varied ($N_t$=20, 50, 200, and 1000).
There are dotted, dashed, dot-dashed, and solid curves at 
$Q^2$=200 GeV$^2$ for $N_t$=20, 50, 200, and 1000 respectively.} 
\label{fig:dqs-nt}
\end{figure}
%%%%%%%%%%%%%%%%%%%%%%%%%%%%%%%% figure %%%%%%%%%%%%%%%%%%%%%%%%%%%%%%%%%%%%%%

\vfill\eject
$\ \ \ $

\vspace{2.5cm}
%%%%%%%%%%%%%%%%%%%%%%%%%%%%%%%% figure %%%%%%%%%%%%%%%%%%%%%%%%%%%%%%%%%%%%%%
\begin{figure}[h]
   \begin{center}
      \epsfig{file=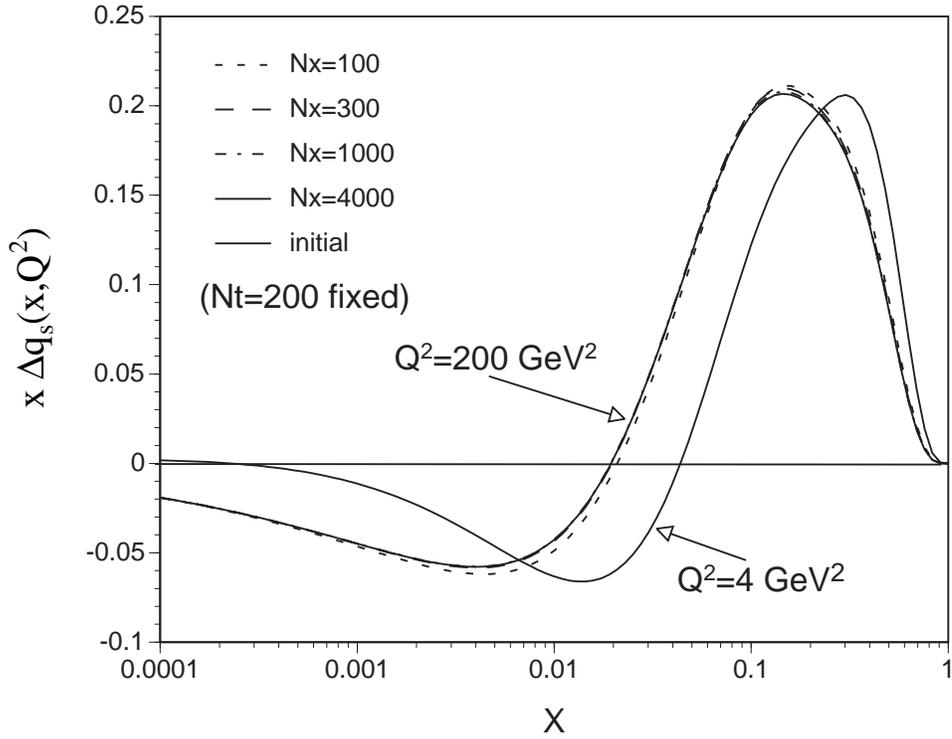,width=15.5cm}
   \end{center}
\caption{$N_x$ dependence of singlet evolution results is shown.
The initial distribution is the same in Fig. \ref{fig:dqs-nt}.
$N_t$=200 is fixed and $N_x$ is varied ($N_x$=100, 300, 1000, and 4000).
There are dotted, dashed, dot-dashed, and solid curves at 
$Q^2$=200 GeV$^2$ for $N_x$=100, 300, 1000, and 4000 respectively.} 
\label{fig:dqs-nx}
\end{figure}
%%%%%%%%%%%%%%%%%%%%%%%%%%%%%%%% figure %%%%%%%%%%%%%%%%%%%%%%%%%%%%%%%%%%%%%%

\vfill\eject
$\ \ \ $

\vspace{2.5cm}
%%%%%%%%%%%%%%%%%%%%%%%%%%%%%%%% figure %%%%%%%%%%%%%%%%%%%%%%%%%%%%%%%%%%%%%%
\begin{figure}[h]
   \begin{center}
      \epsfig{file=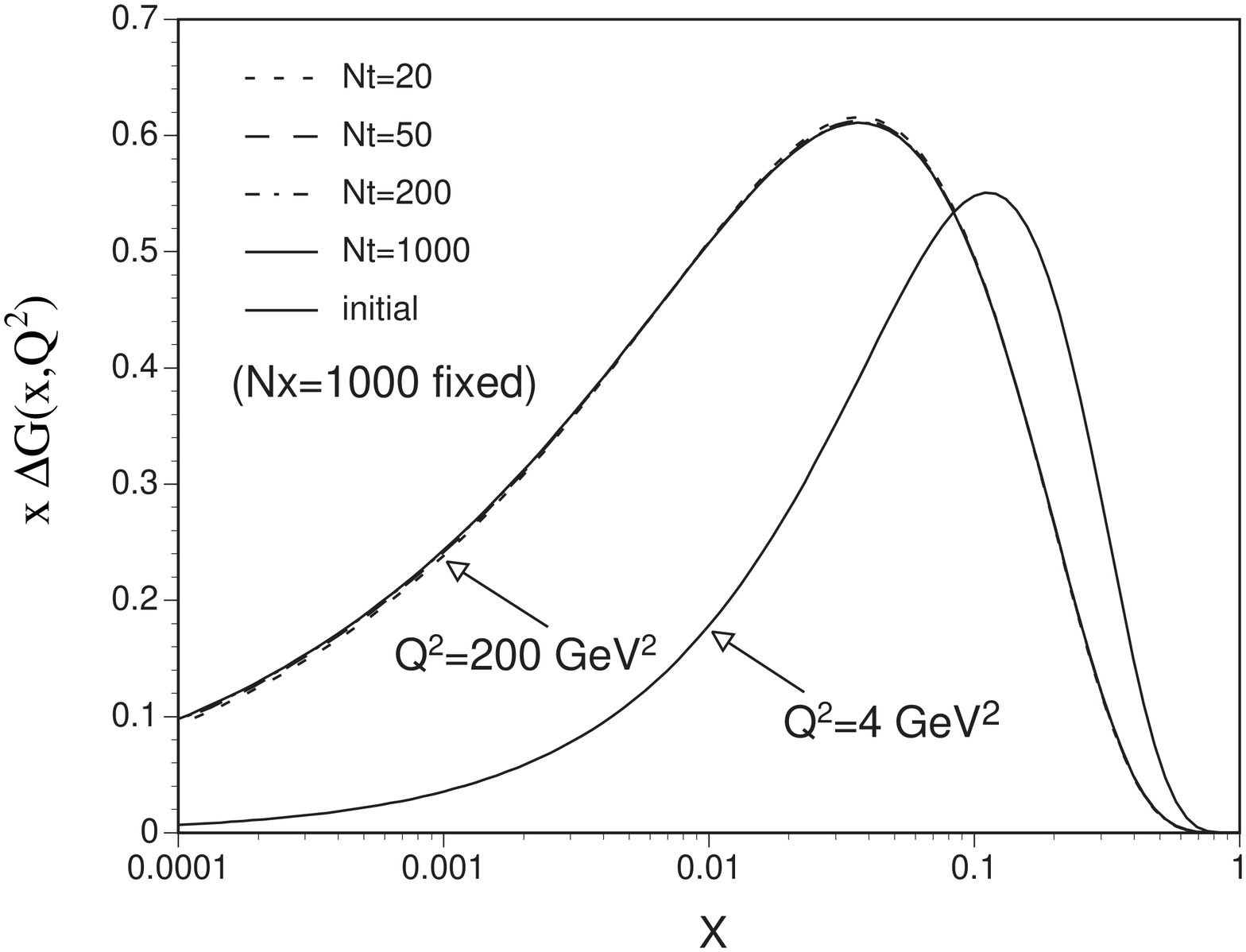,width=15.5cm}
   \end{center}
\caption{$N_t$ dependence of gluon evolution results is shown.
Notations are the same with those in Fig. \ref{fig:dqs-nt}.} 
\label{fig:dg-nt}
\end{figure}
%%%%%%%%%%%%%%%%%%%%%%%%%%%%%%%% figure %%%%%%%%%%%%%%%%%%%%%%%%%%%%%%%%%%%%%%

\vfill\eject
$\ \ \ $

\vspace{2.5cm}
%%%%%%%%%%%%%%%%%%%%%%%%%%%%%%%% figure %%%%%%%%%%%%%%%%%%%%%%%%%%%%%%%%%%%%%%
\begin{figure}[h]
   \begin{center}
      \epsfig{file=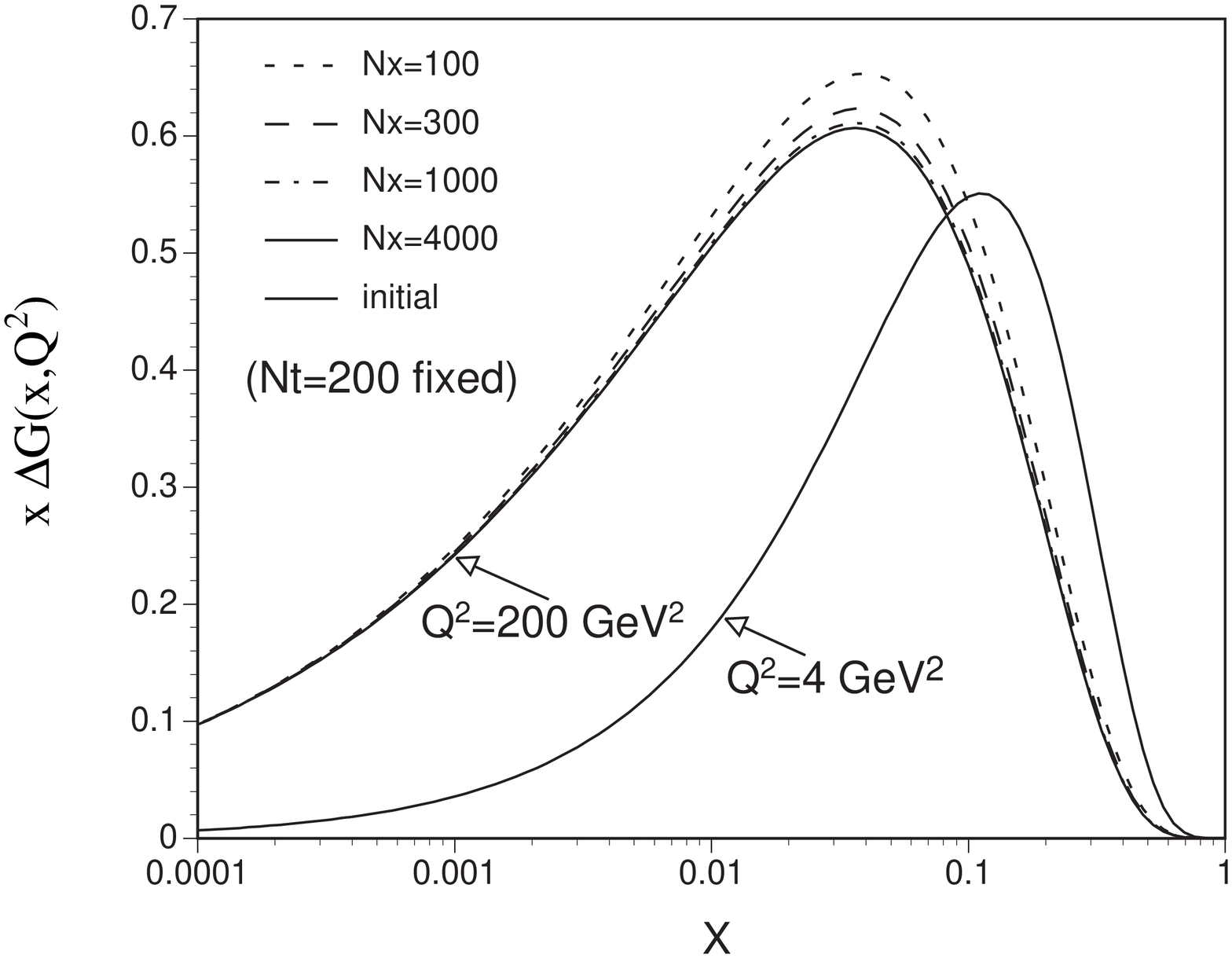,width=15.5cm}
   \end{center}
\caption{$N_x$ dependence of gluon evolution results is shown.
Notations are the same with those in Fig. \ref{fig:dqs-nx}.} 
\label{fig:dg-nx}
\end{figure}
%%%%%%%%%%%%%%%%%%%%%%%%%%%%%%%% figure %%%%%%%%%%%%%%%%%%%%%%%%%%%%%%%%%%%%%%

\vfill\eject
$\ \ \ $

\vspace{2.5cm}
%%%%%%%%%%%%%%%%%%%%%%%%%%%%%%%% figure %%%%%%%%%%%%%%%%%%%%%%%%%%%%%%%%%%%%%%
\begin{figure}[h]
   \begin{center}
      \epsfig{file=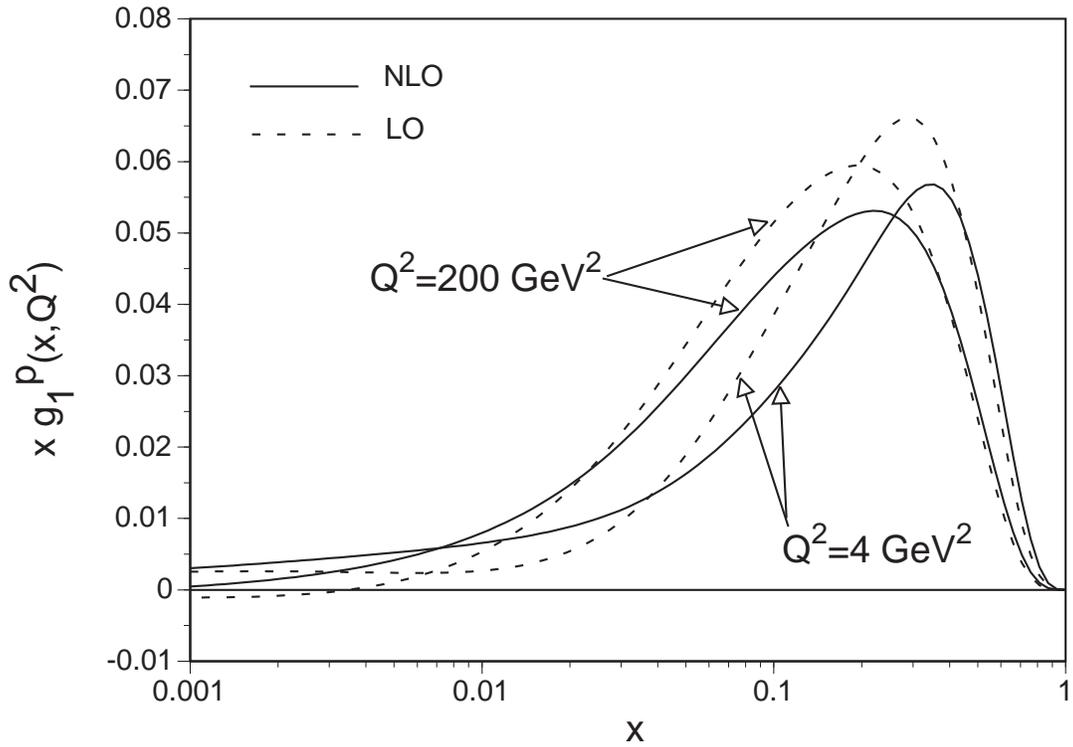,width=15.5cm}
   \end{center}
\caption{$Q^2$ evolution results for the proton structure function $g_1$
are shown. The solid curves are $x g_1$ at $Q^2$=4 and 200 GeV$^2$
in the NLO case, and the dashed ones are in the LO.}
\label{fig:dg1}
\end{figure}
%%%%%%%%%%%%%%%%%%%%%%%%%%%%%%%% figure %%%%%%%%%%%%%%%%%%%%%%%%%%%%%%%%%%%%%%

\vfill\eject
$\ \ \ $

\vspace{2.5cm}
%%%%%%%%%%%%%%%%%%%%%%%%%%%%%%%% figure %%%%%%%%%%%%%%%%%%%%%%%%%%%%%%%%%%%%%%
\begin{figure}[h]
   \begin{center}
      \epsfig{file=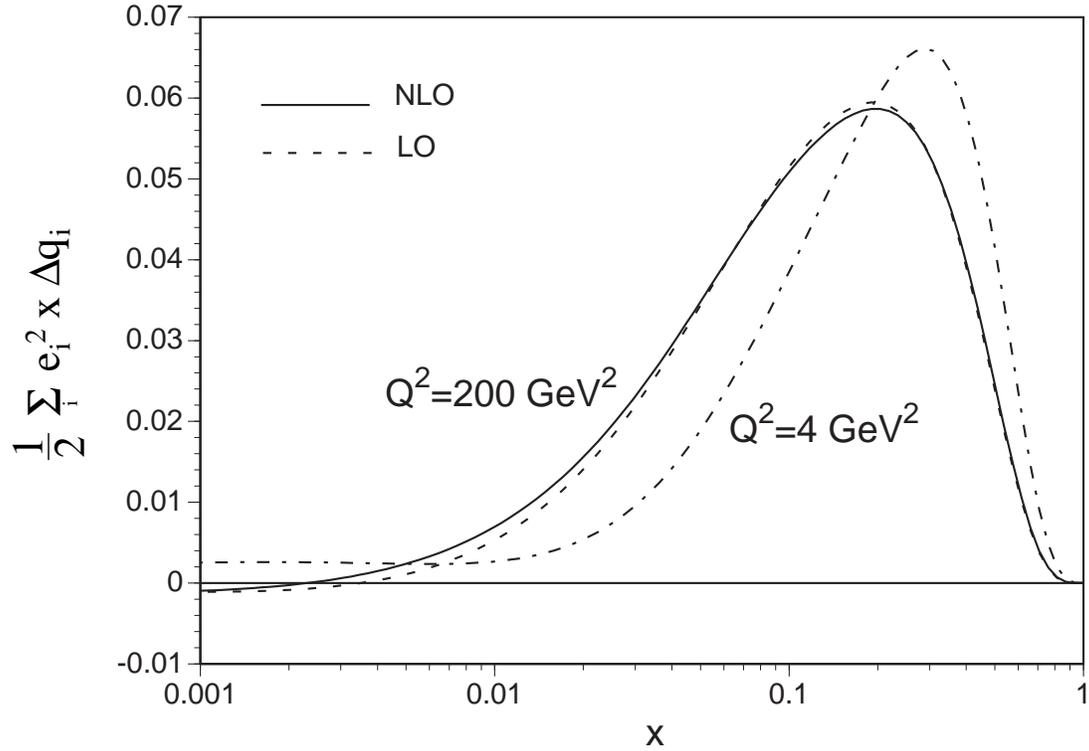,width=15.5cm}
   \end{center}
\caption{$Q^2$ evolution results for the distribution
$\sum_i e_i^2 \, x \, \Delta q_i /2$ are shown.
The same distributions are assumed at $Q^2$=4 GeV$^2$ and 
the evolved distributions are shown by the solid curve 
in the NLO case and by the dashed one in the LO
at 200 GeV$^2$.}
\label{fig:dqg1}
\end{figure}
%%%%%%%%%%%%%%%%%%%%%%%%%%%%%%%% figure %%%%%%%%%%%%%%%%%%%%%%%%%%%%%%%%%%%%%%

\vfill\eject
$\ \ \ $

\vspace{2.5cm}
%%%%%%%%%%%%%%%%%%%%%%%%%%%%%%%% figure %%%%%%%%%%%%%%%%%%%%%%%%%%%%%%%%%%%%%%
\begin{figure}[h]
   \begin{center}
      \epsfig{file=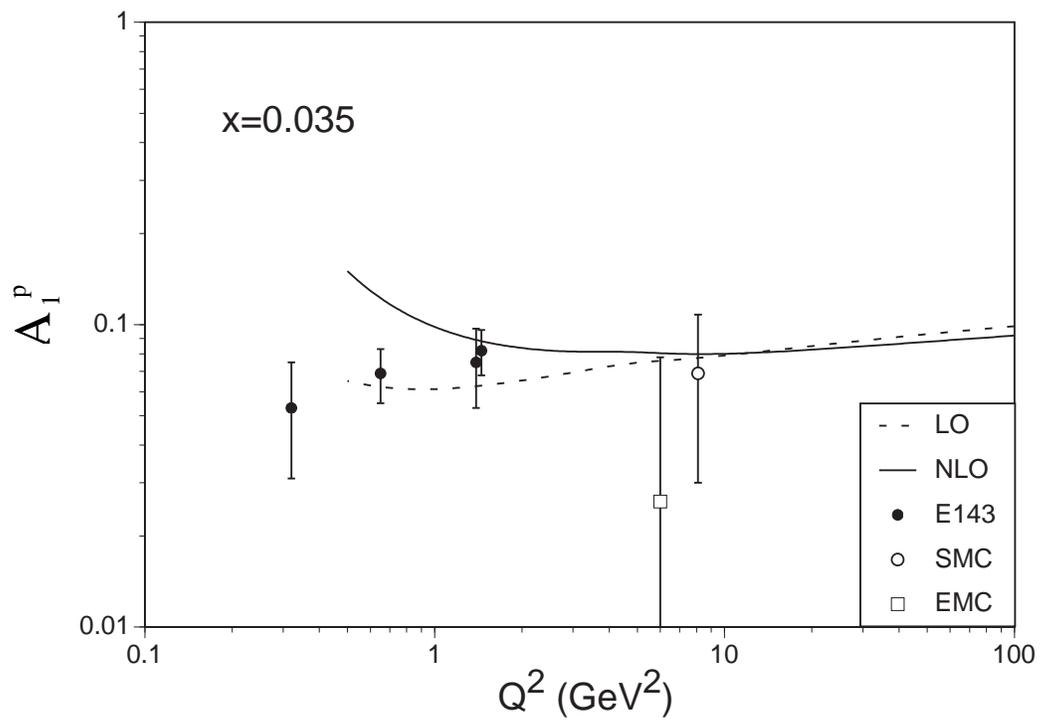,width=15.5cm}
   \end{center}
\caption{$Q^2$ dependence of the spin asymmetry $A_1$ for the proton
is calculated in the LO and NLO cases. Calculated results at $x=0.035$
are compared with SLAC-143 \cite{E143}, 
EMC \cite{EMC}, and SMC \cite{SMC} experimental data.
The solid curve is in the NLO case and the dashed one is in the LO.
} 
\label{fig:a-1}
\end{figure}
%%%%%%%%%%%%%%%%%%%%%%%%%%%%%%%% figure %%%%%%%%%%%%%%%%%%%%%%%%%%%%%%%%%%%%%%

\vfill\eject
$\ \ \ $

\vspace{2.5cm}
%%%%%%%%%%%%%%%%%%%%%%%%%%%%%%%% figure %%%%%%%%%%%%%%%%%%%%%%%%%%%%%%%%%%%%%%
\begin{figure}[h]
   \begin{center}
      \epsfig{file=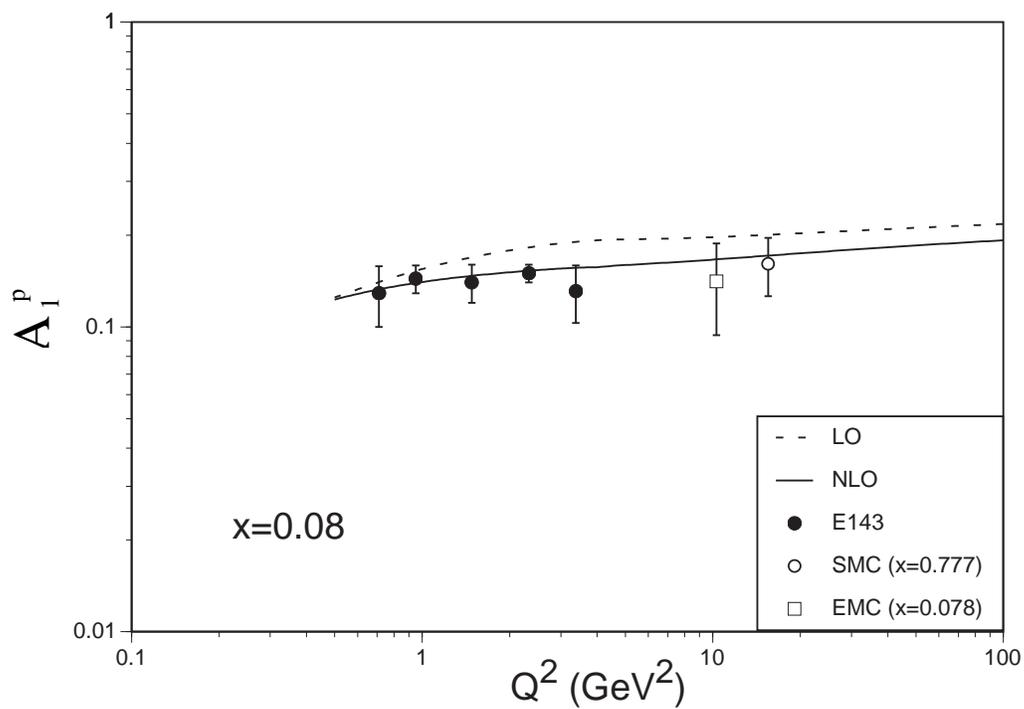,width=15.5cm}
   \end{center}
\caption{$Q^2$ dependence of the spin asymmetry $A_1$ at $x=0.08$.
Notations are same with those in Fig. \ref{fig:a-1}.}
\label{fig:a-2}
\end{figure}
%%%%%%%%%%%%%%%%%%%%%%%%%%%%%%%% figure %%%%%%%%%%%%%%%%%%%%%%%%%%%%%%%%%%%%%%

\vfill\eject
$\ \ \ $

\vspace{2.5cm}
%%%%%%%%%%%%%%%%%%%%%%%%%%%%%%%% figure %%%%%%%%%%%%%%%%%%%%%%%%%%%%%%%%%%%%%%
\begin{figure}[h]
   \begin{center}
      \epsfig{file=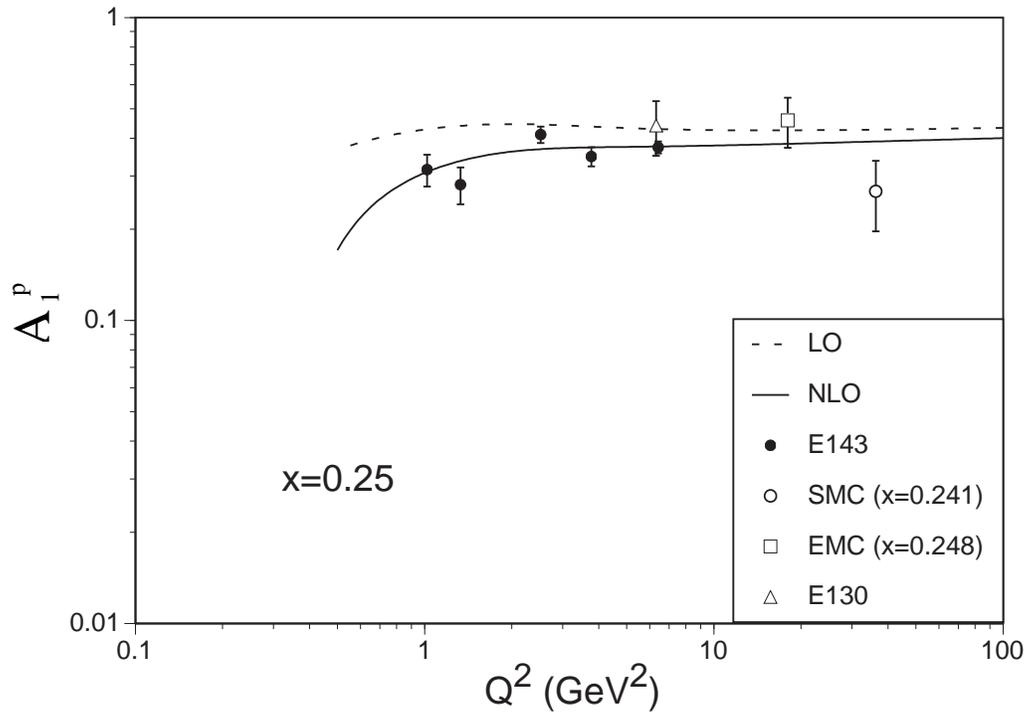,width=15.5cm}
   \end{center}
\caption{$Q^2$ dependence of the spin asymmetry $A_1$ at $x=0.25$.
Notations are same with those in Fig. \ref{fig:a-1}.
The SLAC-E130 data \cite{E130} is added.}
\label{fig:a-3}
\end{figure}
%%%%%%%%%%%%%%%%%%%%%%%%%%%%%%%% figure %%%%%%%%%%%%%%%%%%%%%%%%%%%%%%%%%%%%%%

\end{document}